\documentclass{article}

\usepackage[utf8]{inputenc}
\usepackage{microtype,multibib}
\usepackage{graphicx}
\usepackage{subfigure}
\usepackage{booktabs}
\usepackage{amsmath}
\usepackage{amsfonts}
\usepackage{amsthm}
\usepackage{commath,todonotes}

\usepackage{hyperref}

\usepackage[accepted]{icml2019}

\newcommand{\me}{\mathrm{e}}

\newcommand{\RR}{\mathbb{R}}

\newcommand{\NN}{\mathbb{N}}

\newcommand{\CC}{\mathbb{C}}
\renewcommand{\tilde}{\widetilde}

\usepackage{xcolor,mdframed}

\newenvironment{attn}[1][]{%
   \begin{mdframed}[%
      backgroundcolor={red!15}, hidealllines=true,
      skipabove=0.7\baselineskip, skipbelow=0.7\baselineskip,
      splitbottomskip=2pt, splittopskip=4pt, #1]%
   \makebox[0pt]{
      \smash{
         \fontsize{24pt}{24pt}\selectfont
         \hspace*{-19pt}
         \raisebox{-2pt}{
            {\color{red!70!black}\sffamily\bfseries !}
         }%
      }%
   }%
}{\end{mdframed}}

\begin{document}

\twocolumn[
\icmltitlerunning{Adversarial Generation of Time-Frequency Features}
\icmltitle{Adversarial Generation of Time-Frequency Features\\
\large with application in audio synthesis}



\icmlsetsymbol{equal}{*}

\begin{icmlauthorlist}
\icmlauthor{Andr\'es Marafioti}{ARI}
\icmlauthor{Nicki Holighaus}{ARI}
\icmlauthor{Nathana\"el Perraudin}{SDSC}
\icmlauthor{Piotr Majdak}{ARI}
\end{icmlauthorlist}

\icmlaffiliation{ARI}{Acoustics Research Institute, Austrian Academy of Sciences, Wohllebengasse 12--14, 1040 Vienna, Austria.}
\icmlaffiliation{SDSC}{Swiss Data Science Center, ETH Z\"urich, Universitätstrasse 25, 8006 Z\"urich}

\icmlcorrespondingauthor{Andr\'es Marafioti}{amarafioti@kfs.oeaw.ac.at}

\icmlkeywords{Machine Learning, ICML}

\vskip 0.3in
]



\printAffiliationsAndNotice{}  

\begin{abstract}
Time-frequency (TF) representations provide powerful and intuitive features for the analysis of time series such as audio. But still, generative modeling of audio in the TF domain is a subtle matter. Consequently, neural audio synthesis widely relies on directly modeling the waveform and previous attempts at unconditionally synthesizing audio from neurally generated invertible TF features still struggle to produce audio at satisfying quality. In this article, focusing on the short-time Fourier transform, we discuss the challenges that arise in audio synthesis based on generated invertible TF features and how to overcome them. We demonstrate the potential of deliberate generative TF modeling by training a generative adversarial network (GAN) on short-time Fourier features. We show that by applying our guidelines, our TF-based network was able to outperform a state-of-the-art GAN generating waveforms directly, despite the similar architecture in the two networks. 
\end{abstract}

\section{Introduction}
Despite the recent advance in machine learning and generative modeling, synthesis of natural sounds by neural networks remains a challenge. Recent solutions rely on, among others, classic recurrent neural networks \citep[e.g., SampleRNN,][]{Mehri2016}, dilated convolutions \citep[e.g., WaveNet,][]{Wavenet2016}, and generative adversarial networks~\citep[e.g., WaveGAN,][]{donahue2019wavegan}. Especially, the latter offers a promising approach in terms of flexibility and quality. Generative adversarial networks~\citep[GANs,][]{GoodfellowGAN2014} rely on two competing neural networks trained simultaneously in a two-player min-max game: The generator produces new data from samples of a random variable; The discriminator attempts to distinguish between these generated and real data. During the training, the generator's objective is to fool the discriminator, while the discriminator attempts to learn to better classify real and generated (fake) data. Since their introduction, GANs have been improved in various ways \citep[e.g.,][]{arjovsky2017wasserstein,gulrajani2017improved}. For images, GANs have been used to great success~\cite{karras2017progressive,brock2018large}. For audio, GANs enable the generation of a signal at once even for durations in the range of seconds~\cite{donahue2019wavegan}.

The neural generation of realistic audio remains a challenge, because of its complex structure, with dependencies on various temporal scales. In order to address this issue, a network generating audio is often complemented with another neural network or prior information. For example, the former may require a system of two parallel neural networks \cite{parallel_wavenet2017}, leading overall to more complex systems, while the latter can take the form of a separate conditioning of networks \cite{Tacotron2_2017,sotelo2017char2wav,nsynth2017}. It is usually beneficial to train neural networks on a high-level representation of sound, instead on the time-domain samples. For example, Tacotron 2~\citep{Tacotron2_2017} relies on non-invertible mel-frequency spectrograms. Generation of a time-domain signal from the mel coefficients is then achieved by training a conditioned WaveNet to act as a vocoder.

Time-frequency (TF) domain representations of sound are successfully used in many applications and rely on well-understood theoretical foundations. They  have been widely applied to neural networks, e.g., for solving discriminative tasks, in which they outperform networks directly trained on the waveform \cite{dieleman2014end, Pons2017}. Further, TF representations are used to parameterize neural synthesizers, e.g., Tacotron 2 mentioned above or Timbretron~\cite{huang2018timbretron}, which modifies timbre by remapping constant-Q TF coefficients of sound, conditioning a WaveNet synthesizer.
Despite the success of TF representations for sound \textit{analysis}, why, one could ask, has neural \textit{sound generation} via invertible TF representations only seen limited success?

In fact, there \textit{are} neural networks generating invertible TF representations for sound synthesis. They were designed to perform a specific task such as source separation~\cite{fan2018svsgan,muth2018improving}, speech enhancement~\cite{pascual2017segan}, or audio inpainting~\cite{marafioti2018context} and use a specific and well-chosen setup for TF processing. While the general rules for the parameter choice are not the main focus of those contributions, these rules are highly relevant when it comes to synthesizing sound from a set of TF coefficients generated, e.g., by a neural network.

When both the TF representation and its parameters are appropriately chosen, we generate a highly structured, invertible representation of sound, from which time-domain audio can be obtained using efficient, content-independent reconstruction algorithms. In that case, we do not need to train a problem-specific neural synthesizer. Hence, in this article, we discuss important aspects of neural generation of TF representations particularly for sound synthesis. We focus on the short-time Fourier transform \citep[STFT, e.g.,][]{Allen1977,rawe90}, the best understood and widely used TF representation in the field of audio processing. First, we revisit some properties of the continuous STFT \cite{po76,auger2012phase,gr01} and the progress in phaseless reconstruction of audio signals from STFT coefficients \cite{pruuvsa2017noniterative}. Then, we discuss these properties in the context of the discrete STFT in order to compile guidelines for the choice of STFT parameters ensuring the reliability of sound synthesis and to provide tools monitoring the training progress of the generative models. For the latter, we introduce a novel, experimental measure for the consistency of the STFT. Eventually, we demonstrate the applicability of our guidelines by introducing TiFGAN, a network which generates audio using a TF representation. We provide perceptual and numerical evaluations of TiFGAN demonstrating improved audio quality compared to a state-of-the-art GAN for audio synthesis\footnote{During the preparation of this manuscript, the work \cite{engel2019gansynth} became publicly available. 
In addition to well chosen STFT parameters, usage of the time-direction phase derivative enabled 
their model, GANSynth, to produce significantly better results than previous methods.
The authors kindly provided us with details of their implementation, enabling a preliminary discussion of similarities and differences to our guidelines.}. Our software, complemented by instructive examples, is available at \url{http://tifgan.github.io}.

\begin{figure*}[t!]
    \centering
    \includegraphics[width=0.75\linewidth]{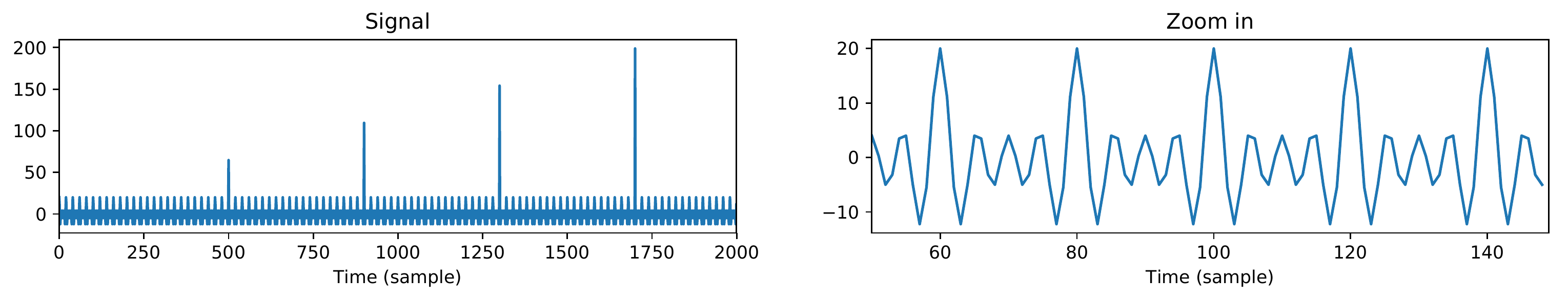}
    \vspace{-0.2cm}
    \includegraphics[width=0.75\linewidth]{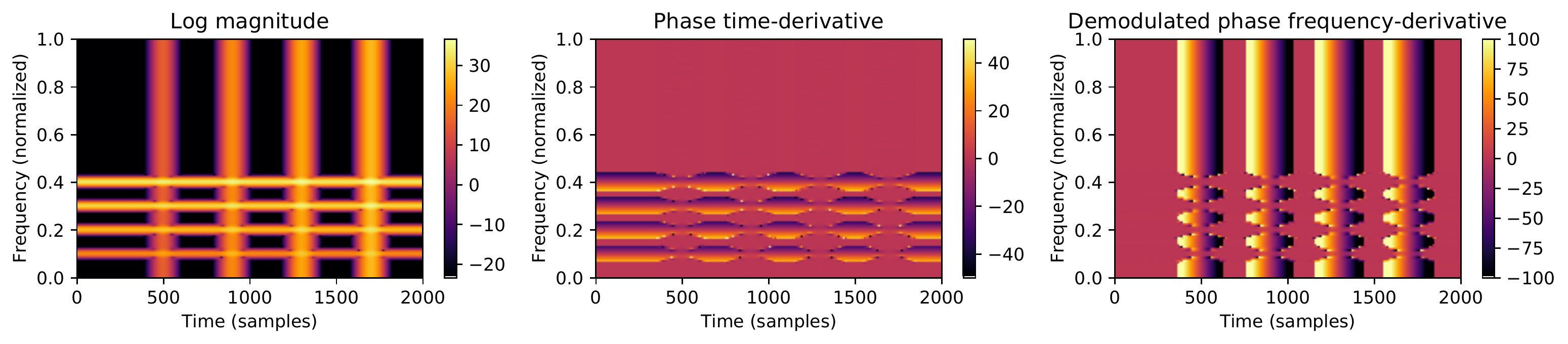}
    \vspace{-0.2cm}
    \caption{Signal representations. Top row: waveform of a test signal (pure tone and pulses). Bottom row: STFT features: log magnitudes (left), time-direction phase derivatives (center) and frequency-direction phase derivatives (right). For small log magnitude, phase derivatives were set to zero. Frequency-direction derivative was computed after demodulation.}
    \label{fig:general_fig}
    \vspace{-0.5cm}
\end{figure*}

\section{Properties of the STFT}\label{sec:STFT}

  The rich structure of the STFT is particularly apparent in the continuous setting of square-integrable functions, i.e. functions in $\mathbf{L}^2(\RR)$. Thus, we first discuss the core issues that arise in the generation of STFTs within that setting, recalling established theory along the way, and then move to discuss these issues in the setting of discrete STFTs.
  
  \subsection{The continuous STFT}\label{sec:contSTFT}
  
  The STFT of the function $f\in\mathbf{L}^2(\RR)$ with respect to the window $\varphi\in\mathbf{L}^2(\RR)$ is given by 
  \begin{equation}\label{eq:STFTcont}
    \operatorname{V}_\varphi f(x,\omega) = \int_\RR f(t)\overline{\varphi(t-x)}\me^{-2\pi i\omega t}\, dt
  \end{equation}
  The variable $(x,\omega)\in\RR^2$ indicates that $\operatorname{V}_\varphi f(x,\omega)$ describes the time-frequency content of $f$ at time $x$ and frequency $\omega$. The STFT is complex-valued and can be rewritten in terms of two real-valued functions as $\operatorname{V}_\varphi f(x,\omega) = \exp(\textrm{M}_\varphi(x,\omega)+i\mathrm{\phi}_\varphi(x,\omega))$, whenever $\operatorname{V}_\varphi f(x,\omega)\neq 0$. The logarithmic magnitude (\emph{log-magnitude}) $\textrm{M}_\varphi$ is uniquely defined, but the \emph{phase} $\mathrm{\phi}_\varphi$ is only defined modulo $2\pi$.
  Further, while $\textrm{M}_\varphi$ is a smooth, slowly varying function, $\mathrm{\phi}_\varphi$ may vary rapidly and is significantly harder to model accurately. Nonetheless, both functions are intimately related. If $\varphi(t) = \varphi_\lambda(t) := e^{-\pi t^2/\lambda}$ is a Gaussian window, this relation can be made explicit~\cite{po76,auger2012phase} through the phase-magnitude relations
  \begin{equation}\label{eq:origderiv1lam}
  \begin{split}
      \pd{\mathrm{\phi}_{\varphi_\lambda}}{x}(x,\omega) &=
    \lambda^{-1}\pd{\textrm{M}_{\varphi_\lambda}}{\omega}(x,\omega),\\
    \pd{\mathrm{\phi}_{\varphi_\lambda}}{\omega}(x,\omega) & = - \lambda\pd{\textrm{M}_{\varphi_\lambda}
    }{x}(x,\omega) - 2\pi x,
    \end{split}
    \end{equation}
    where $\pd{}{\bullet}$ denotes partial derivatives with respect to $\bullet$. Hence, as long as we avoid zeros of $\operatorname{V}_\varphi f$, the phase $\mathrm{\phi}_{\varphi_\lambda}$ can be recovered from $\textrm{M}_{\varphi_\lambda}$ up to a global constant. Since the STFT is invertible, we can recover $f$ from $\textrm{M}_{\varphi_\lambda}$ up to a global phase factor as well, such that it is sufficient to 
    model only the magnitude $\textrm{M}_{\varphi_\lambda}$. 
    
    Note that the partial phase derivatives are of interest by themselves. In contrast to the phase itself, they provide an intuitive interpretation as local instantaneous frequency and time and are useful in various applications~\cite{dolson1986phase,aufl95}. Further, as suggested by \eqref{eq:origderiv1lam}, the phase derivatives might be a more promising modeling target than the phase itself, at least after unwrapping and demodulation\footnote{Formally, demodulation is simply adding $2\pi x$ to $\pd{\mathrm{\phi}_{\varphi_\lambda}}{\omega}(x,\omega)$.} as detailed in \cite{arfib2011time}.
    
    Note that not every function $F\in\mathbf{L}^2(\RR^2)$ is the STFT of a time-domain signal because the STFT operator $\operatorname{V}_\varphi$ maps $\mathbf{L}^2(\RR)$ to a strict subspace of $\mathbf{L}^2(\RR^2)$. Formally, assuming that the window $\varphi$ has unit norm, the inverse STFT is given by the adjoint operator $\operatorname{V}_\varphi^\ast$ of $\operatorname{V}_\varphi$ and we have $\operatorname{V}_\varphi^\ast(\operatorname{V}_\varphi f) = f$ for all $f$. Now, if $F\in \mathbf{L}^2(\RR^2)$ is not in the range of $\operatorname{V}_\varphi$, then $f = \operatorname{V}_\varphi^\ast F$ is a valid time-domain signal, but $F \neq \operatorname{V}_\varphi f$, i.e., $F$ is an \emph{inconsistent} representation of $f$, and the TF structure of $F$ will be distorted in $\operatorname{V}_\varphi f$. 
    
    In the presence of phase information, consistency of $F$ can be evaluated simply by computing the norm difference $\|F-\operatorname{V}_\varphi(\operatorname{V}_\varphi^\ast F)\|$ which can also serve as part of a training objective. 
    If only magnitudes $\widetilde{\textrm{M}}$
    are available, we can theoretically exploit the phase-magnitude relations \eqref{eq:origderiv1lam}, reconstruct the phase, and then evaluate consistency. 
    Unless otherwise specified, coefficients that are not necessarily consistent are indicated by the symbol $\sim$, e.g., generated magnitudes $\widetilde{\textrm{M}}$. 
    In practice, phase recovery from the magnitude $\widetilde{\textrm{M}}$ introduces errors of its own and the combined process may become too expensive to be attractive as a training objective. Thus, it might be preferable to evaluate consistency of the generated magnitude 
    directly, which, for Gaussian windows, can be derived from \eqref{eq:origderiv1lam}
    \begin{equation}\label{eq:consistency_contlam}
     \left(\lambda\tfrac{\partial^2}{\partial x^2} + \lambda^{-1}\tfrac{\partial^2}{\partial \omega^2}\right)\textrm{M}_{\varphi_\lambda}(x,\omega) = - 2\pi.
  \end{equation} 
   
  Note that, although \cite{po76} already observed that $\widetilde{\textrm{M}}$ is an STFT magnitude if and only if \eqref{eq:consistency_contlam} holds (and $e^{\widetilde{\textrm{M}}}$ is square-integrable), our contribution is, to our knowledge, the first to exploit this relation to evaluate consistency. 
  Furthermore, the phase-magnitude relations \eqref{eq:origderiv1lam} and the consistency equivalence 
  \eqref{eq:consistency_contlam} 
  can be traced back to the relation of Gaussian STFTs to a certain space of analytic 
  functions~\cite{bargman1961,complexanalysis78}.
  
  In the context of neural networks, the ultimate goal of the generation process is to obtain a time-domain signal, but we can only generate a finite number of STFT coefficients. Therefore,  it is essential that inversion from the generated values is possible and synthesis of the time-domain signal is robust to distortions. In mathematical terms, this requires a window function $\varphi$ and time and frequency steps $a,b\in\RR^+$ specifying a snug STFT (or Gabor) \emph{frame}~\cite{ch16}. While a comprehensive discussion of STFT frames is beyond the scope of this article, it is generally advisable to match $a,b$ to the width of $\varphi$ and its Fourier transform $\hat{\varphi}$. In the case where both $\varphi$ and $\hat{\varphi}$ are at least remotely bell-shaped, a straightforward measure of their widths are the standard deviations $\sigma_\varphi = \sigma(\varphi/\|\varphi\|_{\mathbf L^1})$ and $\sigma_{\hat{\varphi}} = \sigma(\hat{\varphi}/\|\hat{\varphi}\|_{\mathbf L^1})$. Hence, we expect good results if $a/b = \sigma_\varphi/\sigma_{\hat{\varphi}}$. For Gaussian windows $\varphi_\lambda$, we have $\sigma_{\varphi_\lambda}/\sigma_{\widehat{\varphi_\lambda}} = \lambda$, such that $\lambda$ is often referred to as \emph{time-frequency ratio}. For such $\varphi_\lambda$, the choice $a/b = \lambda$ is conjectured to be optimal\footnote{In the sense of the frame bound ratio, which is a measure of transform stability~\cite{ch16}.} in general~\cite{best03}, and proven to be so for $(ab)^{-1}\in\NN$~\cite{faulhuber2017optimal}. Furthermore, the relations  \eqref{eq:origderiv1lam} and \eqref{eq:consistency_contlam} only hold exactly for the undecimated STFT and must be approximated. For this approximation to be reliable, $ab$ must be small enough. The theory suggests that $ab\leq 1/4$ is generally required for reliable reconstruction of signals from the magnitude alone ~\cite{balan2006signal}. For larger $ab$, the values of the STFT become increasingly independent and little exploitable (or learnable) structure remains. 
   
   These considerations provide useful guidelines for the choice of STFT parameters. In the following, we translate them into a discrete implementation.
   
   \subsection{The discrete STFT}\label{sec:discSTFT}
   
   The STFT of a finite, real signal $s\in\RR^L$, with the \emph{analysis window} $g\in\RR^L$,
   time step $a\in \NN$ and $M \in\NN$ frequency channels is given by 
   \begin{equation}\label{eq:STFT_final}
   \begin{split}
      \operatorname{S}_g(s)[m,n] & = \sum_{l\in \underline{L}} s[l]g[l-na]e^{-2\pi i ml/M},
   \end{split}
   \end{equation}
   for
   $n\in \underline{N}, m\in\underline{M}$, where we denote, for any $j\in\NN$, $\underline{j} = [0,\ldots,j-1]$ and indices are to be understood modulo $L$. Similar to the continuous case, we can write $\operatorname{S}_g(s)[m,n] = \exp(\textrm{M}_g[m,n] + i\mathrm{\phi}_g[m,n])$, with log-magnitude $\textrm{M}_g$ and phase $\mathrm{\phi}_g$. 
   The vectors $\operatorname{S}_g(s)[\cdot,n]\in\CC^N$ and $\operatorname{S}_g(s)[m,\cdot]\in\CC^{M}$ are called the $n$-th (time) segment 
  and $m$-th (frequency) channel of the STFT, respectively. 
  
  Let $b=L/M$. Then, the ratio $M/a = L/(ab)$ is a measure of the transform redundancy and the STFT is overcomplete (or redundant) if $M/a > 1$. If $s$ and $g$ are real-valued, all time segments are conjugate symmetric and it is sufficient to store the first
  $M_{\RR} = \lfloor M/2\rfloor$ channels only, such that the STFT matrix can be reduced to the size $M_{\RR} \times N$. 
  
  The inverse STFT with respect to the \emph{synthesis window} $\tilde{g}\in\RR^L$ can be written as 
  $\tilde{s}[l] = \sum_{n\in\underline{N}} \sum_{m\in\underline{M}}  \operatorname{S}_g(s)[m,n]\tilde{g}[l-na]e^{2\pi iml/M}$, $l\in\underline{L}$. We say that 
  $\tilde{g}$ is a \emph{dual window} for $g$, if $\tilde{s} = s$ for all $s\in\RR^L$ \cite{st98-1,janssen1997continuous,rawe90}.
  
  Note that the number of channels $M$ can be smaller than the number of nonzero samples $L_g$ of $g$, as long as 
  $a$ and $M$ respect the widths of $g$ and its Fourier transform $\hat{g}$ as discussed in Sec. \ref{sec:contSTFT}. This yields $aM \approx L\frac{\sigma_g}{\sigma_{\hat{g}}}$ as a general guideline
  with $\sigma_g = \sigma(g/\|g\|_{\ell^1})$ and $\sigma_{\hat{g}} = \sigma(\hat{g}/\|\hat{g}\|_{\ell^1})$.
  Furthermore, with the redundancy $M/a \geq 4$, there is sufficient dependency between the values of the STFT, e.g., to facilitate magnitude-only reconstruction. In our experience, this choice represents a lower bound for reliability of discrete approximation of \eqref{eq:origderiv1lam}.

  \begin{attn}
  Implementations of STFT in  SciPy and Tensorflow introduce a phase skew dependent on the (stored) window length $L_g$ (usually $L_g\ll L$) and with severe effects on any phase analysis and processing if not accounted for. This can be addressed with the conversion between \eqref{eq:STFT_final} and other conventions presented in the supplementary material~\ref{sec:supp-conventions} and \cite{arfib2011time,ltfatnote042}.
  \end{attn} 
  
  \subsection{Phase recovery and the phase-magnitude relationship}\label{ssec:phaserec}
 
  \begin{figure}[t]
    \centering
    \includegraphics[scale=0.35]{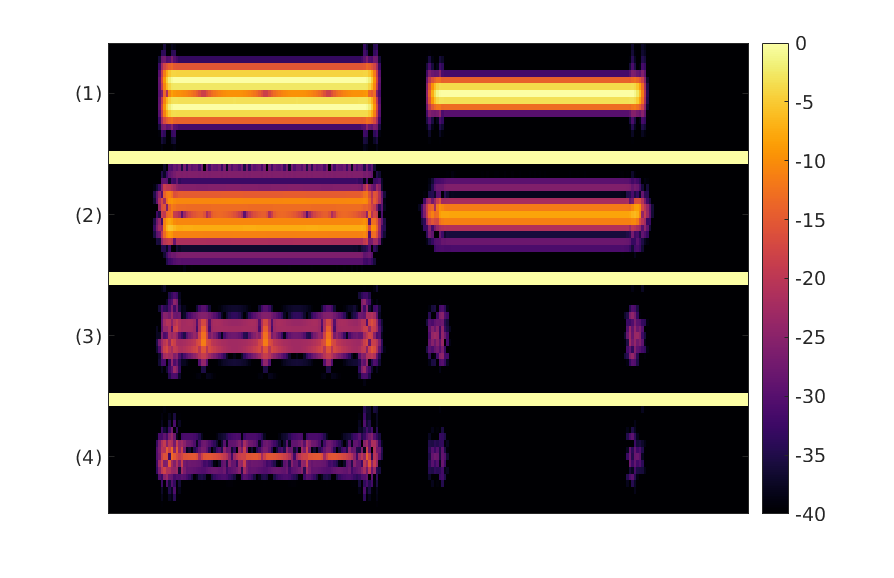}
    \vspace{-16pt}
    \caption{Overview of spectral changes resulting from different phase reconstruction methods. (1) Original log-magnitude, (2-4) log-magnitude differences between original and signals restored with (2) cumulative sum along channels (initialized with zeros), (3) PGHI from phase derivatives (4) PGHI from magnitude only and phase estimated from Eq. \eqref{eq:phase-magnitude}.}
    \label{fig:phaserecs}
    \vspace{-24pt}
  \end{figure}
  
  Let $\partial_\bullet$ denote some discrete partial differentiation scheme. Discrete approximation of the phase-magnitude relationship \eqref{eq:origderiv1lam} results in 
  \begin{equation}\label{eq:phase-magnitude}
    \begin{split}
    \partial_n \mathrm{\phi}_g[m,n] & \approx \frac{aM}{\lambda} \partial_m \textrm{M}_g[m,n],\\
    \partial_m \mathrm{\phi}_g[m,n] & \approx -\frac{\lambda}{aM} \partial_n \textrm{M}_g[m,n] - 2\pi na/M,
    \end{split}
  \end{equation}
  as derived in \cite{pruuvsa2017noniterative}. For non-Gaussian windows $g$, choosing $\lambda = \sigma_g / \sigma_{\hat{g}}$ has shown surprisingly reliable results, but accuracy of \eqref{eq:phase-magnitude} depends on the proximity of the window $g$ to a Gaussian nonetheless. Optimal results are obtained for Gaussian windows at redundancy $M/a = L$. While STFTs with $M/a = 4$ perform decently and are considered in our network architecture. In Section \ref{ssec:consistencyEval} we show that a further, moderate increase in redundancy has the potential to further elevate synthesis quality.
  As an alternative to estimating the phase derivatives from the magnitude, it may be feasible to generate estimates of the phase derivative directly within a generative model.
  
  It may seem straightforward to restore the phase from its time-direction derivative by summation along frequency channels as proposed in \cite{engel2019gansynth}. Even on real, unmodified STFTs, the resulting phase misalignment introduces cancellation between frequency bands resulting in energy loss, see Fig.~\ref{fig:phaserecs}(2) for a simple example. In practice, such cancellations often leads to clearly perceptible changes of timbre\footnote{See \url{http://tifgan.github.io} for examples.}. Moreover, in areas of small STFT magnitude, the phase is known to be unreliable~\cite{balazs2016pole} and sensitive to distortions~\cite{alaifari2019stability,alaifari2017phase,mallat2015phase}, such that it cannot be reliably modelled and synthesis from generated phase derivatives is likely to introduce more distortion. 
  Phase-gradient heap integration~\citep[PGHI,][]{pruuvsa2017noniterative} relies on the phase-magnitude relations \eqref{eq:phase-magnitude} and bypasses phase instabilities by avoiding integration through areas of small magnitude, leading to significantly better and more robust phase estimates $\widetilde{\mathrm{\phi}}$, see Fig.~\ref{fig:phaserecs}(4). PGHI  often outperforms more expensive, iterative schemes relying on alternate projection, e.g., Griffin-Lim \cite{griffin1984signal,le2010fast,Perraudin2013}, at the phaseless reconstruction (PLR) task. Generally, PLR relies heavily on consistent STFT magnitude for good results. Note that the integration step in PGHI can also be applied if phase derivative estimates from some other source are available, e.g., when training a network to learn time- and frequency-direction phase derivatives. For an example, see Fig. \ref{fig:phaserecs}(3).
  
\subsection{Consistency of the STFT}\label{ssec:consistency}

  The space of valid STFTs with a given window is a lower dimensional subspace of all complex-valued matrices of size $M_{\RR} \times N$ and a given, generated matrix $\widetilde{\textrm{S}}$ may be very far from the STFT of any time-domain signal, even if it \emph{looks} correct. To prevent artifacts, it is important to ensure that $\widetilde{\textrm{S}}$ is \emph{consistent}. Let $\operatorname{iS}_{\tilde{g}}$ denote the inverse STFT with the dual window $\tilde{g}$, see Sec. \ref{sec:discSTFT}. Consistency of $\widetilde{\textrm{S}}$ can be evaluated by computing the \emph{projection error}
  \begin{equation}\label{eq:projection}
    \operatorname{e}^\textrm{proj} = \|\widetilde{\textrm{S}}- \operatorname{S}_g(\operatorname{iS}_{\tilde{g}}(\widetilde{\textrm{S}}))\|, 
  \end{equation}
  where $\|\cdot\|$ denotes the Euclidean norm. When $\operatorname{e}^\textrm{proj}$ is large, its effects on the synthesized signal are unpredictable and degraded synthesis quality must be expected. Although $\operatorname{e}^\textrm{proj}$ is an accurate consistency measure, it can be computationally expensive. Further, its use for evaluating the consistency of magnitude-only data is limited: When preceded by phase recovery, $\operatorname{e}^\textrm{proj}$ is unable to distinguish the error introduced by the employed PLR method from inconsistency of the provided magnitude data. 
  
  As an alternative, we instead propose an experimental measure that evaluates consistency of the log-magnitude directly. The proposed consistency measure exploits the consistency relation \eqref{eq:consistency_contlam}. An approximation in the spirit of \eqref{eq:phase-magnitude} yields
  \begin{equation}\label{eq:laplacian}
    \frac{\lambda}{a^2}\partial_n^2 \textrm{M}_g[m,n] + \frac{M^2}{\lambda} \partial_m^2 \textrm{M}_g[m,n] \approx -2\pi. 
  \end{equation}
  In practice, and in particular at moderate redundancy, we found \eqref{eq:laplacian} to be prone to approximation errors. Experimentally, however, a measure inspired by the sample Pearson correlation~\cite{lyons1991practical} provided promising results. 
  Let $\tilde{\textrm{M}}$ be the generated magnitude, we have
  \begin{equation}
     \textrm{DM}_n = |\partial_n^2 \tilde{\textrm{M}} +  \tfrac{\pi a^2}{\lambda}|,\ 
     \textrm{DM}_m = |\partial_m^2 \tilde{\textrm{M}} + \tfrac{\pi \lambda}{M^2}|,
  \end{equation}
   where the terms $\pi a^2/\lambda$ and $\pi \lambda/M^2$ are obtained by distributing the shift $2\pi$ in \eqref{eq:laplacian} equally to both terms on the left hand side. We define the consistency $\varrho(\tilde{\textrm{M}})$ of $\tilde{\textrm{M}}$ as
  \begin{equation}\label{eq:logM-consistency}
    \varrho(\tilde{\textrm{M}}) := \operatorname{r}(\textrm{DM}_n,\textrm{DM}_m),
  \end{equation}
  where $\operatorname{r}(X,Y)$ is the sample Pearson correlation coefficient of the paired sets of samples $(X,Y)$. If the equality is satisfied in \eqref{eq:laplacian}, then  $\varrho(\tilde{\textrm{M}})=1$. Conversely if $\varrho(\tilde{\textrm{M}})\approx 0$, then \eqref{eq:laplacian} is surely violated and the representation is inconsistent.  The performance of $\varrho$ as consistency measure is discussed in Section \ref{ssec:consistencyEval} below.
  
  \section{Performance of the consistency measure}\label{ssec:consistencyEval}
  
  The purpose of the consistency measure $\varrho$ is to determine whether a generated log-magnitude is likely to be close to the log-magnitude STFT of a signal, i.e. it is consistent. As discussed above, consistency is crucial to prevent degraded synthesis quality. Hence, it is important to evaluate the dependence of its properties on changes in the redundancy, the window function and its sensitivity to distortion. 
  
\begin{figure}[t]
      \centering
      \includegraphics[scale=0.4]{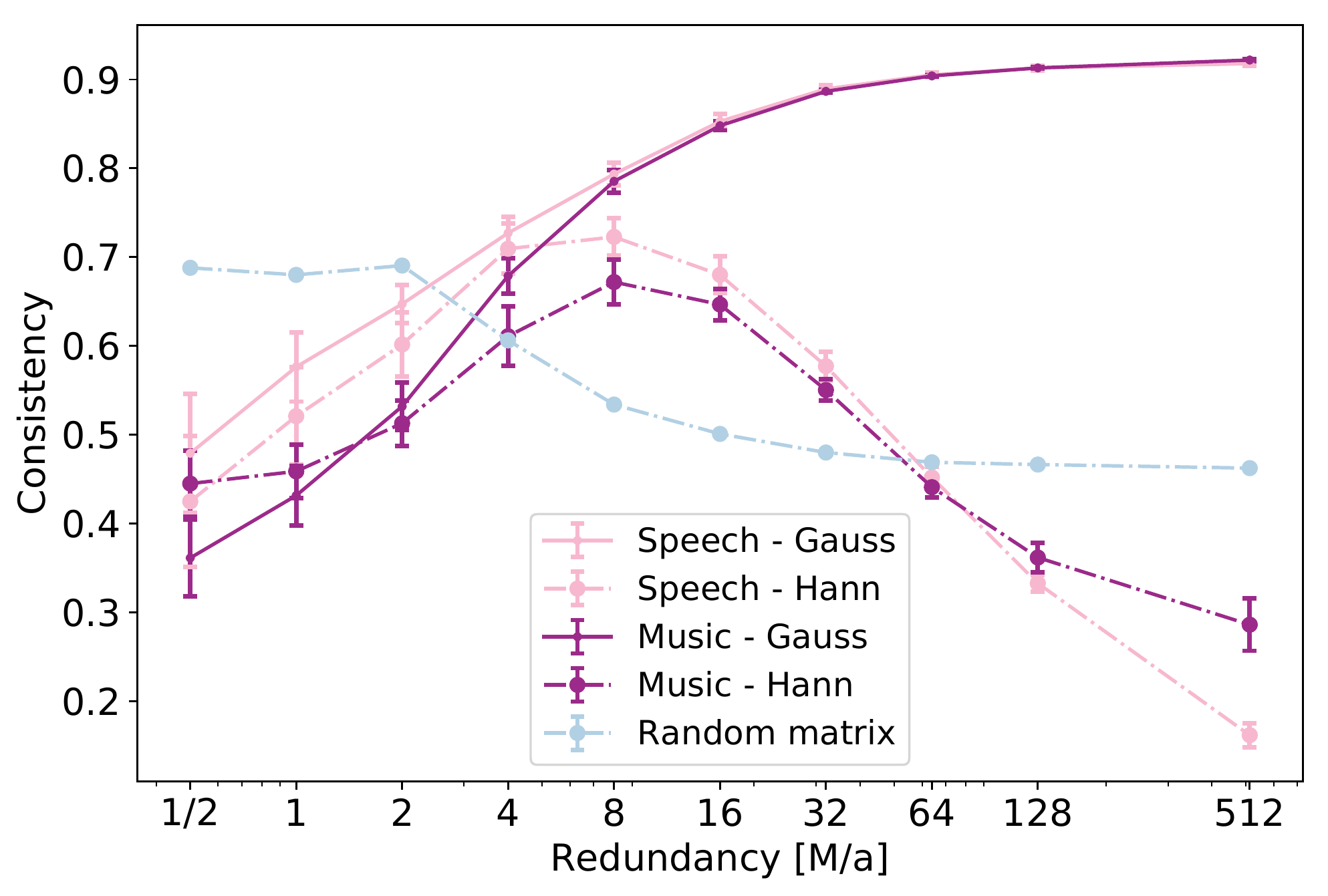}
      \vspace{-0.5cm}
      \caption{Consistency as function of the redundancy for various time-domain windows. Random matrix from Gaussian distribution.}
      \label{fig:consistency_per_redundancy}
    \vspace{-18pt}
\end{figure}

  In a first test, we compute the mean and standard deviation of $\varrho$ on a speech and a music dataset, see Section \ref{sec:TiFGAN} for details, at various redundancies, using Gaussian and Hann windows with time-frequency ratio $\lambda \approx 4$ and STFT parameters satisfying $aM/L = 4$, see Fig. \ref{fig:consistency_per_redundancy}. We note that a Gaussian random matrix takes surprisingly high values for $\varrho$ and, thus, $\varrho$ is not reliable below redundancy $4$. For Gaussian windows, mean consistency increases with redundancy, while the standard deviation decreases, indicating that $\varrho$ becomes increasingly reliable and insensitive to signal content. This analysis suggests that a redundancy of $8$ or $16$ could lead to notable improvements. At redundancy $4$, spectrograms for both types of data score reliably better than the random case, with speech scoring higher than music. The Hann window scores worse than the Gaussian on average in all conditions, with a drop in performance above $M/a = 16$. This indicates that $\varrho$ is only suitable to evaluate consistency of  Hann window log-magnitudes for redundancies in the range $6$ to $16$.

\begin{figure}[th!]
      \centering
      \includegraphics[scale=0.4]{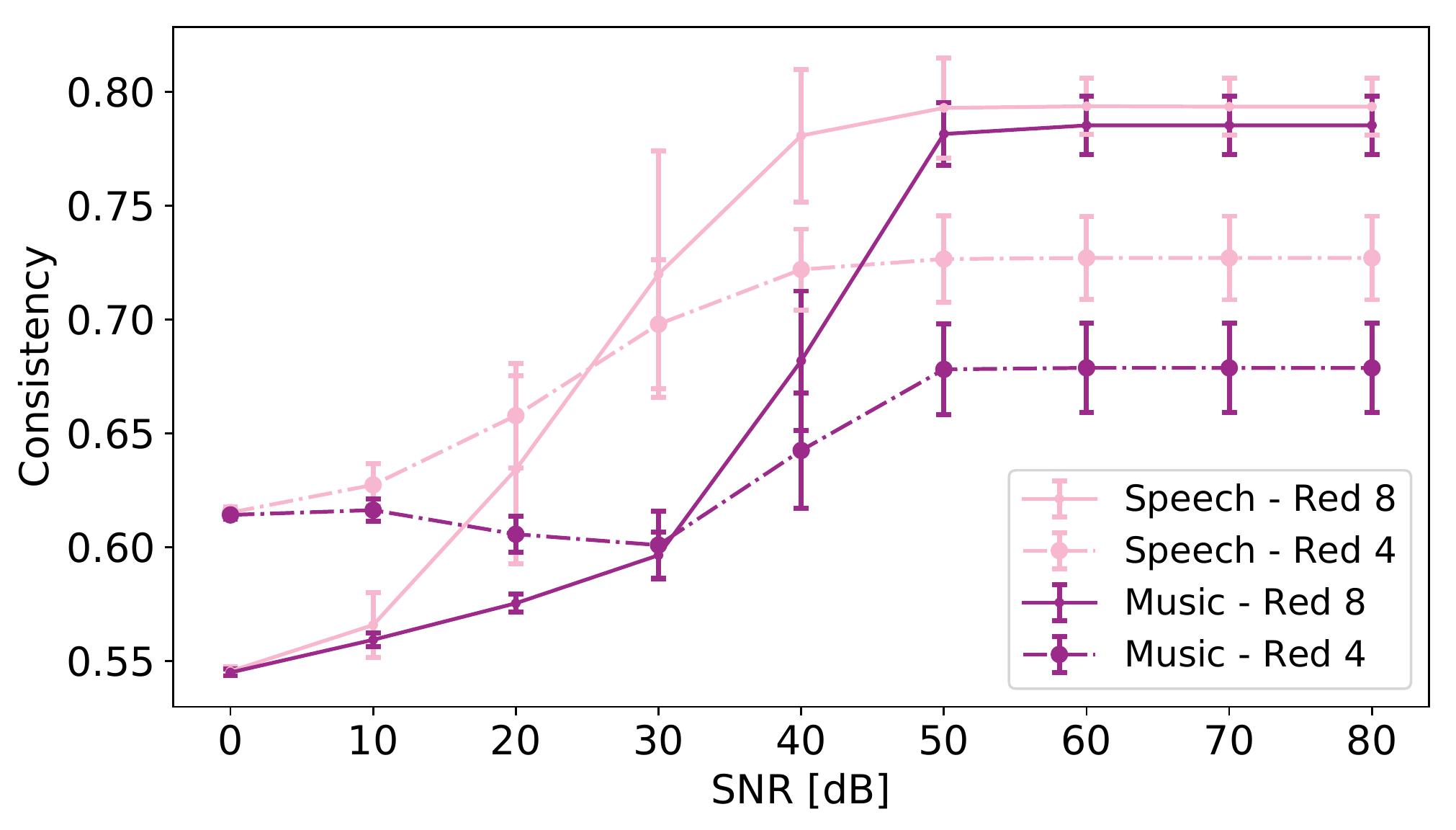}
      \vspace{-0.8cm}
      \caption{Consistency as a function of SNR obtained by adding complex-valued Gaussian noise to the STFT coefficients.}
      \label{fig:consistency_noisy_nofloor}
      \vspace{-0.4cm}
  \end{figure}

In a second test, we fix a Gaussian STFT with redundancies $4$ and $8$ and evaluate the behaviour of $\varrho$ under deviations from true STFT magnitudes. To this end, we add various levels of uniform Gaussian noise to the STFT before computing the log--magnitude, see Fig. \ref{fig:consistency_noisy_nofloor}. At redundancy $8$ we observe a monotonous decrease of consistency with increasing noise level. In fact, the consistency converges to the level of random noise at high noise levels. Especially for music, $\varrho$ is sensitive to even small levels of noise. At redundancy $4$, the changes are not quite as pronounced, but the general trend is similar. While this is not a full analysis of the measure $\varrho$, it is reasonable to expect that models that match the value of $\varrho$ closely generate approximately consistent log-magnitudes

Furthermore, the results suggest that $\varrho$ has a low standard deviation across data of the same distribution. In the context of GANs, where no obvious convergence criterion applies, $\varrho$ can thus assist the determination of convergence and divergence by tracking 
  \begin{equation}\label{eq:rel_consistency}
      \gamma = \left|\mathbb{E}_{\bf{M} \sim \mathbb{P}_{M_\text{real}}}\left[ \varrho(\bf{M}) \right] - \mathbb{E}_{\bf{M} \sim \mathbb{P}_{M_\text{fake}}}\left[ \varrho(\bf{M}) \right]\right|.
  \end{equation} 

  \section{Time-Frequency Generative Adversarial Network (TiFGAN)}
\label{sec:TiFGAN}
  To demonstrate the potential of the guidelines and principles for generating short-time Fourier data presented in Section \ref{sec:STFT}, we apply them to TiFGAN, which unconditionally generates audio using a TF representation and improves on the current state-of-the-art for audio synthesis with GANs. 
  For the purpose of this contribution, we restrict to generating $1$ second of audio, or more precisely $L=16384$ samples sampled at $16$\,kHz. For the short-time Fourier transform, we fix the minimal redundancy that we consider reliable, i.e., $M/a = 4$ and select $a=128, M=512$, such that $M_\RR = 257$, $N = L/a = 128$ and the STFT matrix $S$ is of size $\CC^{M_\RR\times N}$. This implies that the frequency step is $b = L/M = 32$, such that we chose for the analysis window $g$ a (sampled) Gaussian with time-frequency ratio $\lambda = 4 = aM/L$. Since the Nyquist frequency is not expected to hold significant information for the considered signals, we drop it to arrive at a representation size of $256\times 128$, which is well suited to processing using strided convolutions. 

\begin{figure}[ht!]
    \centering
    \includegraphics[scale=0.12]{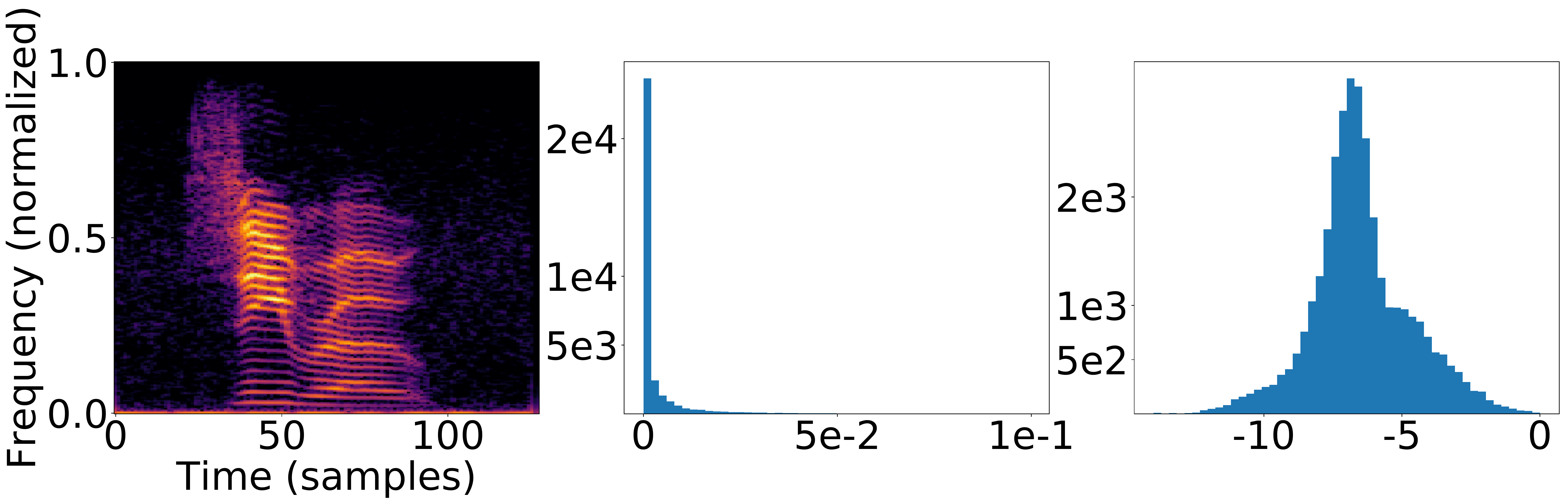}
    \vspace{-2em}
    \caption{From left to right: log-magnitude spectrogram, distribution of the magnitude, distribution of the log-magnitude.}
    \label{fig:distrib}
\end{figure}

The log-magnitude distribution is closer to human sound perception and, as show in Fig.~\ref{fig:distrib}, it doesn't have the large tail of the magnitude STFT coefficients, therefore we use it for the training data. To do so, we first normalize the STFT magnitude to have maximum value $1$, such that the log-magnitude is confined in $(-\infty,0]$. Then, the dynamic range of the log-magnitude is limited by clipping at $-r$ (in our experiments $r=10$), before scaling and shifting to the range of the generator output $[-1,1]$, i.e. dividing by $r/2$ before adding constant $1$.
The network trained to generate log-magnitudes will be referred to as TiFGAN-M. Generation of, and synthesis from, the log-magnitude STFT is the main focus of this contribution. Nonetheless, we also trained a variant architecture TiFGAN-MTF for which
we additionally provided the time- and frequency-direction derivatives of the (unwrapped, demodulated) phase\footnote{Phase derivatives were obtained using the \texttt{gabphasegrad} function in the Large Time-Frequency Analysis Toolbox \citep[LTFAT,][]{ltfatnote030}.}~\cite{arfib2011time,dolson1986phase}. 

For TiFGAN-M, the phase derivatives are estimated from the generated log-magnitude following \eqref{eq:phase-magnitude}. For both TiFGAN-M and TiFGAN-MTF, the phase is reconstructed from the phase derivative estimates using phase-gradient heap integration~\citep[PGHI,][]{pruuvsa2017noniterative}, which requires no iteration, such that reconstruction time is comparable to simply integrating the phase derivatives.
For synthesis from the STFT, we use the \emph{canonical dual window}~\cite{st98-1,ch16}, precomputed using the Large Time-Frequency Analysis Toolbox \citep[LTFAT,][]{ltfatnote030}, avalable at \url{ltfat.github.io}.

\begin{figure}[t!]
    \centering
    \vspace{-0.3cm}
    \includegraphics[width=1\columnwidth]{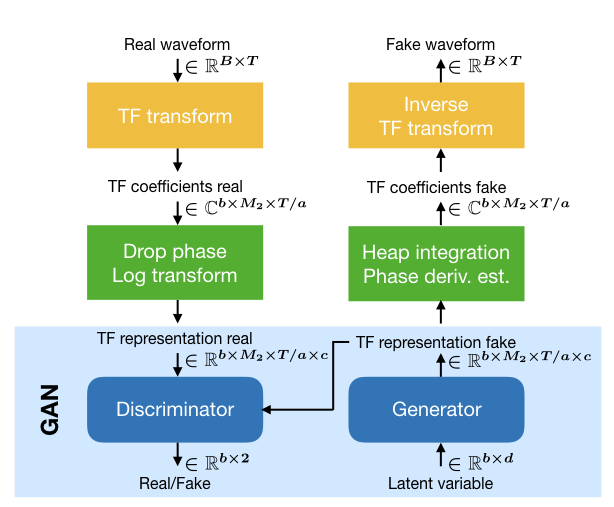}
    \vspace{-.4cm}
    \caption{The general architecture with parameters $T = 16384$, $a = 128$, $M_{2} = 256$ $c=1,3$, $d=100$. Here  $b = 64$ is the batch size. The orange and green steps describe the pre- and post-processing stages.}
    \vspace{-0.5cm}
    \label{fig:general_architecture}
\end{figure}

\paragraph{GAN architecture:} 
The TiFGAN architecture, depicted in Fig. \ref{fig:general_architecture}, is an adaptation of DCGAN \citep{radford2015unsupervised} and similarly to WaveGAN and SpecGAN \cite{donahue2019wavegan}, we add one convolutional layer each to generator and discriminator to enable the generation of larger matrices. Moreover, we generate data of size $(256,128)$, a rectangular array of twice the width and four times the height of DCGANs output, and twice the height of SpecGAN, such that we also adapted the filter shapes to better reflect and capture the rectangular shape of the training data\footnote{When training on piano data, we also observed that, when using square filters, the frequency content of note onsets was unnaturally dispersed over time. This effect was notably reduced after switching to tall filters}. Precisely in comparison to SpecGAN, we use filters of shape $(12, 3)$ instead of the $~31\%$ smaller $(5,5)$. To compensate, we further reduce the number of filter channels of the fully-connected layer and the first convolutional layer of the generator by a factor of $2$. Since these two layers comprise the majority of parameters, our architecture only has $10\%$ more parameters than SpecGAN in total. More details on the architecture can be found in Section \ref{sec:supp-gandetails} of the supplementary material. 

\begin{figure}[h]
    \centering
    \includegraphics[scale=0.3]{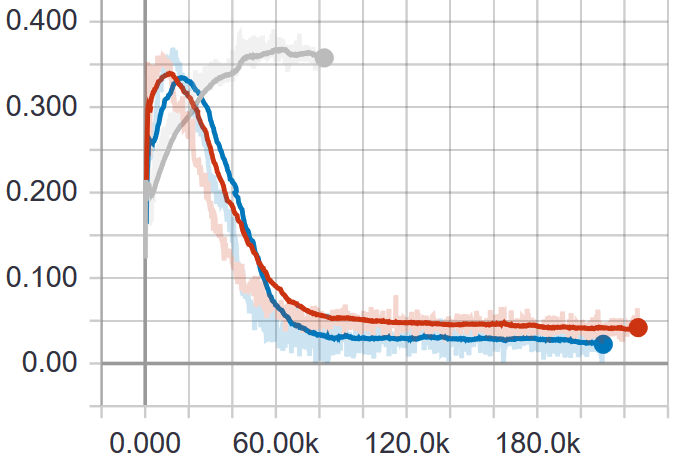}
    \caption{Eq.~\eqref{eq:rel_consistency} for three networks. Gray: failed network. Red: TiFGAN-M. Blue: TiFGAN-MTF as in Sec.~\ref{sec:TiFGAN}.}
    \label{fig:consistency_as_convergence_2}
    \vspace{-0.5cm}
\end{figure}

\paragraph{Training:} 
During training of TiFGAN, we monitored the relative consistency $\gamma$  of the generator \eqref{eq:rel_consistency} in addition to the adversarial loss, negative critic and gradient penalty. In the optimization phase, networks that failed to train well, could often be detected to diverge in consistency and discarded after less than $50k$ steps of training ($1$ day), while promising candidates quickly started to converge towards the consistency of the training data, i.e., $\gamma \rightarrow 0$, see Fig.~\ref{fig:consistency_as_convergence_2}. Networks with smaller $\gamma$ synthesized better audio, but when trained for many steps, they were sometimes less reliable in terms of semantic audio content, e.g., for speech they were more likely to produce gibberish words than with shorter training. Our networks were trained for 200k steps as this seemed to provide reasonably good results in both semantic and audio quality. We optimized the Wasserstein loss \cite{gulrajani2017improved} with the gradient penalty hyperparameter set to $10$ using the ADAM optimizer \cite{Kingma2014} with $\alpha = 10^{-4},\ \beta_1 = 0.5 ,\ \beta_2 = 0.9$ and performed $5$ updates of the discriminator for every update of the generator. For the reference condition, we used the pre-trained WaveGAN network provided by \cite{donahue2019wavegan}\footnote{\url{https://github.com/chrisdonahue/wavegan}}.


\paragraph{Comparison to SpecGAN \cite{donahue2019wavegan}:} TiFGAN is purposefully designed to be similar to SpecGAN\footnote{Note that SpecGAN is of equal size as WaveGAN.} to emphasize that the main cause for improved results is the handling of time-frequency data according to the guidelines in Section \ref{sec:discSTFT}. SpecGAN relies on an STFT of redundancy $M/a = 2$ with Hann window of length $L_g = 256$, time step $a=128$ and $M=256$ channels. According to Section \ref{sec:discSTFT}, this setup is not very well suited to generative modeling. PLR in particular is expected to be unreliable, which is evidenced by the results reported in \cite{donahue2019wavegan}, which employ the classical Griffin-Lim algorithm \cite{griffin1984signal} for PLR. The choice of STFT parameters for SpecGAN fixes a target size of shape $(128,128)$, while for TiFGAN the target size is $(256,128)$. This required some changes to the network architecture, as presented above. Finally, SpecGAN performs a normalization per frequency channel over the entire dataset, preventing the network to learn the natural relations between channels in the STFT log-magnitude, which are crucial for consistency, as shown in Section \ref{sec:discSTFT}. 

\begin{table*}[th!]
\centering
\vspace{-0.1cm}
\begin{tabular}{lccccccc}
\hline
            &  vs TiFGAN-M & vs TiFGAN-MTF &  vs WaveGAN & Cons & RSPE (dB) & IS  & FID  \\ \hline
Real        &  86\% & 90\% & 94\% & 0.70 & -22.0* & 7.98   & 0.5       \\
TiFGAN-M     &  --   & 67\% & 75\% & 0.67 & -13.8 & 5.97   &  26.7      \\
TiFGAN-MTF   &  33\% & --   & 55\% & 0.68 & -12.5*  & 4.48   &  32.6      \\
WaveGAN     &  25\% & 45\% & --   & --   & --    & 4.64   & 41.6\\
\end{tabular}
\vspace{-0.3cm}
\caption{Results of the evaluation. First three left columns: Preference (in \%) of the condition shown in a row over the conditions show in a column, obtained from listening tests. Cons: averaged consistency measure $\rho$. RSPE: as in Eq. \eqref{eq:RSPE}. IS: inception score. FID:  Fr\'echet inception distance. *These values were obtained by discarding the phase and reconstructing from the magnitude only. For the listening tests, the signals contained the full representation.}
\label{tab:comparison}
\vspace{-0.4cm}
\end{table*}

\subsection{Evaluation}\label{ssec:eval}
%

To evaluate the performance of TiFGAN, we trained TiFGAN-M and TiFGAN-MTF using the procedure outlined above on two datasets from \cite{donahue2019wavegan}: (a) Speech, a subset of spoken digits "zero" through "nine" (sc09) from the "Speech Commands Dataset"~\cite{warden2018speech}. This dataset is not curated, some samples are noisy or poorly labeled, the considered subset consists of approximately 23,000 samples. (b) Music, a dataset of $~25$ minutes of piano recordings of Bach compositions, segmented into approximately 19,000 overlapping samples of $1$~s duration.

\paragraph{Evaluation metrics:} For speech and music, we provide audio examples online\footnote{\url{http://tifgan.github.io}}. For speech, we performed listening tests and evaluated the inception score (IS)~\cite{salimans2016improved} and Fr\'echet inception distance~(FID)~\cite{heusel2017gans}, using the pre-trained classifier provided with~\cite{donahue2019wavegan}. For the real data and both variants of TiFGAN, we moreover computed the consistency $\varrho$, see Eq. \eqref{eq:logM-consistency}, and the relative spectral projection error (RSPE) in dB, after phase reconstruction from the log-magnitude, i.e.,
\begin{equation}\label{eq:RSPE}
   10\log_{10}\left( \frac{\||\widetilde{\textrm{S}}|-|\operatorname{S}_g(\operatorname{iS}_{\tilde{g}}(\widetilde{\textrm{S}}))|\|}{\|\widetilde{\textrm{S}}\|}\right),
\end{equation}
where $|\widetilde{\textrm{S}}| = |\operatorname{S}_g(s)|$ in the case of real data and $|\widetilde{\textrm{S}}| = \exp(\widetilde{\textrm{M}})$, with the generated log-magnitude $\widetilde{\textrm{M}}$, for the generated data. 
Phase-gradient heap integration was applied to obtain $\widetilde{\textrm{S}}$ from $|\widetilde{\textrm{S}}|$ (and generated phase derivatives in the case of TiFGAN-MTF). 

Listening tests were performed in a sound booth and sounds were presented via headphones, see supplementary material \ref{sec:supp-booths}. The task involved pairwise comparison of preference between four conditions: real data extracted from the dataset, TiFGAN-M generated examples, TiFGAN-MTF generated examples, and WaveGAN generated examples. In each trial, listeners were provided with two sounds from two different conditions. The question to the listener was "which sound do you prefer?". Signals were selected at random from $600$ pre-generated examples per condition. Each of the six possible combinations was repeated $80$ times in random order, yielding 480 trials per listener. The test lasted approximately 45 minutes including breaks which subjects were allowed to take at any time. Seven subjects were tested and none of them were the authors. A post-screening showed that one subject was not able to distinguish between any of the conditions and thus was removed from the test, yielding in 2880 valid preferences in total from six subjects.

\paragraph{Results:} The results are summarized in Table~\ref{tab:comparison}. On average, the subjects preferred the real samples over WaveGAN's in 94\% of the examples given. For TiFGAN-MTF, the preference decreased to 90\% and for TiFGAN-M further to 86\%.
The large gap between generated and real data can be explained by the experimental setup that enables a very critical evaluation. Nonetheless, it is apparent that TiFGAN-M performed best in the direct comparison to real data by a significant margin. 
Comparison of the other pairings leads to a similar conclusion: Subjects preferred TiFGAN-MTF over WaveGAN in 55\% of the examples given, TiFGAN-M over WaveGAN in 75\% and TiFGAN-M over TiFGAN-MTF in 67\%. While TiFGAN-M clearly outperformed the other networks, TiFGAN-MTF was only slightly more often preferred over WaveGAN. 

The analysis of IS and FID leads to similar conclusions: TiFGAN-M showed a large improvement on both measures over the other conditions, with still a large gap to the real-data performance. On the other hand, comparing WaveGAN to TiFGAN-MTF, the results for both measures are mixed.

When evaluating the magnitude spectrograms generated by TiFGAN-M, TiFGAN-MTF, and those obtained from the real data, we notice that their consistencies are similarly close. Going a step further and applying PGHI to these magnitude spectrograms, the relative projection errors (RSPE) of the two networks are similar, but worse than those of the real signals, meaning that there is room for improvement in this regard. For the listening tests, PGHI was applied to the output of TiFGAN-MTF using the generated phase derivatives. In this setting, the RSPE was -7.5~dB, a substantially smaller value. This confirms our finding that phase reconstruction provides better results than phase generation by our network.

In summary, TiFGAN-M provided a substantial improvement over the previous state-of-the-art in unsupervised adversarial audio generation. Although the results for TiFGAN-MTF are not as clear, we believe that direct generation of phase could provide results on par or better than the magnitude alone and should be systematically investigated.
Further research will focus on avoiding discrepancies between the phase derivatives and the log-magnitude.


\section{Conclusions}  

In this contribution, we considered adversarial generation of a well understood time-frequency representation, namely the STFT. We proposed steps to overcome the difficulties that arise when generating audio in the short-time Fourier domain, taking inspiration from properties of the continuous STFT \cite{po76,auger2012phase,gr01} and from the recent progress in phaseless reconstruction \cite{pruuvsa2017noniterative}. We provided guidelines for the choice of STFT parameters that ensure the reliability of phaseless reconstruction. 
Further, we introduced a new  measure assessing the quality of a magnitude STFT, i.e., the consistency measure. It is computationally cheap and can be used to a-priori estimate the potential success of phaseless reconstruction. In the context of GANs, it can ease the assessment of convergence at training time. 

Eventually, we demonstrated the value of our guidelines in the context of unsupervised audio synthesis with GANs. We introduced TiFGAN, a GAN directly generating invertible STFT representations. Our TiFGANs, trained on speech and music outperformed the state-of-the-art time-domain GAN both in terms of psychoacoustic and numeric evaluation, demonstrating the potential of TF representations in generative modeling. 

In the future, further extensions of the proposed approach are planned towards TF representations on logarithmic and perceptual frequency scales~\cite{brown1991calculation,brown1992efficient,holighaus2013framework,holighaus2019char,necciari18}.

\newpage

\section*{Acknowledgments}
This work has been supported by Austrian Science Fund (FWF) project MERLIN (Modern methods for the restoration of lost information in digital signals;I 3067-N30). We gratefully acknowledge the support of NVIDIA Corporation with the donation of the Titan X Pascal GPU used for this research.
We would like to thank the authors of \cite{engel2019gansynth} for providing us with details of their implementations prior to presenting it at ICLR 2019, allowing us to have their approach as a comparison. We would also like to thank the anonymous reviewers and Peter Balazs for their tremendously helpful comments and suggestions.

\bibliography{gansbib}

\begin{thebibliography}{57}
\providecommand{\natexlab}[1]{#1}
\providecommand{\url}[1]{\texttt{#1}}
\expandafter\ifx\csname urlstyle\endcsname\relax
  \providecommand{\doi}[1]{doi: #1}\else
  \providecommand{\doi}{doi: \begingroup \urlstyle{rm}\Url}\fi

\bibitem[Alaifari \& Grohs(2017)Alaifari and Grohs]{alaifari2017phase}
Alaifari, R. and Grohs, P.
\newblock Phase retrieval in the general setting of continuous frames for
  banach spaces.
\newblock \emph{SIAM journal on mathematical analysis}, 49\penalty0
  (3):\penalty0 1895--1911, 2017.

\bibitem[Alaifari \& Wellershoff(2019)Alaifari and
  Wellershoff]{alaifari2019stability}
Alaifari, R. and Wellershoff, M.
\newblock Stability estimates for phase retrieval from discrete gabor
  measurements.
\newblock \emph{arXiv preprint arXiv:1901.05296}, 2019.

\bibitem[Allen(1977)]{Allen1977}
Allen, J.
\newblock Short term spectral analysis, synthesis, and modification by discrete
  fourier transform.
\newblock \emph{IEEE Transactions on Acoustics, Speech, and Signal Processing},
  25\penalty0 (3):\penalty0 235--238, 1977.
\newblock \doi{10.1109/TASSP.1977.1162950}.

\bibitem[Arfib et~al.(2011)Arfib, Keiler, Z{\"o}lzer, Verfaille, and
  Bonada]{arfib2011time}
Arfib, D., Keiler, F., Z{\"o}lzer, U., Verfaille, V., and Bonada, J.
\newblock Time-frequency processing.
\newblock \emph{DAFX: Digital Audio Effects}, pp.\  219--278, 2011.

\bibitem[Arjovsky et~al.(2017)Arjovsky, Chintala, and
  Bottou]{arjovsky2017wasserstein}
Arjovsky, M., Chintala, S., and Bottou, L.
\newblock Wasserstein generative adversarial networks.
\newblock In \emph{Proc. of ICML}, pp.\  214--223, 2017.

\bibitem[{A}uger \& {F}landrin(1995){A}uger and {F}landrin]{aufl95}
{A}uger, F. and {F}landrin, P.
\newblock {I}mproving the readability of time-frequency and time-scale
  representations by the reassignment method.
\newblock \emph{{I}{E}{E}{E} {T}rans. {S}ignal {P}roc.}, 43\penalty0
  (5):\penalty0 1068 --1089,, may 1995.

\bibitem[Auger et~al.(2012)Auger, Chassande-Mottin, and
  Flandrin]{auger2012phase}
Auger, F., Chassande-Mottin, {\'E}., and Flandrin, P.
\newblock On phase-magnitude relationships in the short-time fourier transform.
\newblock \emph{IEEE Signal Process. Lett.}, 19\penalty0 (5):\penalty0
  267--270, 2012.

\bibitem[Balan et~al.(2006)Balan, Casazza, and Edidin]{balan2006signal}
Balan, R., Casazza, P., and Edidin, D.
\newblock On signal reconstruction without phase.
\newblock \emph{Applied and Computational Harmonic Analysis}, 20\penalty0
  (3):\penalty0 345--356, 2006.

\bibitem[Balazs et~al.(2016)Balazs, Bayer, Jaillet, and
  S{\o}ndergaard]{balazs2016pole}
Balazs, P., Bayer, D., Jaillet, F., and S{\o}ndergaard, P.
\newblock The pole behavior of the phase derivative of the short-time fourier
  transform.
\newblock \emph{Applied and Computational Harmonic Analysis}, 40\penalty0
  (3):\penalty0 610--621, 2016.

\bibitem[Bargmann(1961)]{bargman1961}
Bargmann, V.
\newblock On a {H}ilbert space of analytic functions and an associated integral
  transform part i.
\newblock \emph{Communications on Pure and Applied Mathematics}, 14\penalty0
  (3):\penalty0 187--214, 1961.
\newblock \doi{10.1002/cpa.3160140303}.
\newblock URL
  \url{https://onlinelibrary.wiley.com/doi/abs/10.1002/cpa.3160140303}.

\bibitem[Brock et~al.(2019)Brock, Donahue, and Simonyan]{brock2018large}
Brock, A., Donahue, J., and Simonyan, K.
\newblock Large scale {GAN} training for high fidelity natural image synthesis.
\newblock In \emph{Proc. of ICLR}, 2019.

\bibitem[Brown(1991)]{brown1991calculation}
Brown, J.~C.
\newblock Calculation of a constant {Q} spectral transform.
\newblock \emph{The Journal of the Acoustical Society of America}, 89\penalty0
  (1):\penalty0 425--434, 1991.

\bibitem[Brown \& Puckette(1992)Brown and Puckette]{brown1992efficient}
Brown, J.~C. and Puckette, M.~S.
\newblock An efficient algorithm for the calculation of a constant {Q}
  transform.
\newblock \emph{The Journal of the Acoustical Society of America}, 92\penalty0
  (5):\penalty0 2698--2701, 1992.

\bibitem[{C}hristensen(2016)]{ch16}
{C}hristensen, O.
\newblock \emph{{A}n {I}ntroduction to {F}rames and {R}iesz {B}ases}.
\newblock {A}pplied and {N}umerical {H}armonic {A}nalysis. {B}irkh{\"a}user
  {B}asel, {S}econd edition, 2016.
\newblock ISBN 978-3-319-25611-5; 978-3-319-25613-9.

\bibitem[Conway(1973)]{complexanalysis78}
Conway, J.~B.
\newblock \emph{Functions of one complex variable}.
\newblock Springer-Verlag New York [New York], 1973.
\newblock ISBN 3540900624 0387900616 0387900624.

\bibitem[Dieleman \& Schrauwen(2014)Dieleman and Schrauwen]{dieleman2014end}
Dieleman, S. and Schrauwen, B.
\newblock End-to-end learning for music audio.
\newblock In \emph{Acoustics, Speech and Signal Processing (ICASSP), 2014 IEEE
  International Conference on}, pp.\  6964--6968. IEEE, 2014.

\bibitem[Dolson(1986)]{dolson1986phase}
Dolson, M.
\newblock The phase vocoder: A tutorial.
\newblock \emph{Computer Music Journal}, 10\penalty0 (4):\penalty0 14--27,
  1986.

\bibitem[Donahue et~al.(2019)Donahue, McAuley, and
  Puckette]{donahue2019wavegan}
Donahue, C., McAuley, J., and Puckette, M.
\newblock Adversarial audio synthesis.
\newblock In \emph{Proc. of ICLR}, 2019.

\bibitem[Engel et~al.(2017)Engel, Resnick, Roberts, Dieleman, Norouzi, Eck, and
  Simonyan]{nsynth2017}
Engel, J., Resnick, C., Roberts, A., Dieleman, S., Norouzi, M., Eck, D., and
  Simonyan, K.
\newblock Neural audio synthesis of musical notes with wavenet autoencoders.
\newblock In \emph{Proc. of ICML}, pp.\  1068--1077, 2017.

\bibitem[Engel et~al.(2019)Engel, Agrawal, Chen, Gulrajani, Donahue, and
  Roberts]{engel2019gansynth}
Engel, J., Agrawal, K.~K., Chen, S., Gulrajani, I., Donahue, C., and Roberts,
  A.
\newblock {GANS}ynth: Adversarial neural audio synthesis.
\newblock In \emph{Proc. of ICLR}, 2019.

\bibitem[Fan et~al.(2018)Fan, Lai, and Jang]{fan2018svsgan}
Fan, Z.-C., Lai, Y.-L., and Jang, J.-S.~R.
\newblock {SVSGAN}: Singing voice separation via generative adversarial
  network.
\newblock In \emph{2018 IEEE International Conference on Acoustics, Speech and
  Signal Processing (ICASSP)}, pp.\  726--730. IEEE, 2018.

\bibitem[Faulhuber \& Steinerberger(2017)Faulhuber and
  Steinerberger]{faulhuber2017optimal}
Faulhuber, M. and Steinerberger, S.
\newblock Optimal gabor frame bounds for separable lattices and estimates for
  jacobi theta functions.
\newblock \emph{Journal of Mathematical Analysis and Applications},
  445\penalty0 (1):\penalty0 407--422, 2017.

\bibitem[Goodfellow et~al.(2014)Goodfellow, Pouget-Abadie, Mirza, Xu,
  Warde-Farley, Ozair, Courville, and Bengio]{GoodfellowGAN2014}
Goodfellow, I., Pouget-Abadie, J., Mirza, M., Xu, B., Warde-Farley, D., Ozair,
  S., Courville, A., and Bengio, Y.
\newblock Generative adversarial nets.
\newblock In \emph{Advances in neural information processing systems}, pp.\
  2672--2680, 2014.

\bibitem[Griffin \& Lim(1984)Griffin and Lim]{griffin1984signal}
Griffin, D. and Lim, J.
\newblock Signal estimation from modified short-time fourier transform.
\newblock \emph{IEEE Transactions on Acoustics, Speech and Signal Processing},
  32\penalty0 (2):\penalty0 236--243, 1984.

\bibitem[{G}r{\"o}chenig(2001)]{gr01}
{G}r{\"o}chenig, K.
\newblock \emph{{F}oundations of {T}ime-{F}requency {A}nalysis}.
\newblock {A}ppl. {N}umer. {H}armon. {A}nal. {B}irkh{\"a}user, 2001.

\bibitem[Gulrajani et~al.(2017)Gulrajani, Ahmed, Arjovsky, Dumoulin, and
  Courville]{gulrajani2017improved}
Gulrajani, I., Ahmed, F., Arjovsky, M., Dumoulin, V., and Courville, A.~C.
\newblock Improved training of wasserstein {GAN}s.
\newblock In \emph{Advances in Neural Information Processing Systems}, pp.\
  5767--5777, 2017.

\bibitem[Heusel et~al.(2017)Heusel, Ramsauer, Unterthiner, Nessler, and
  Hochreiter]{heusel2017gans}
Heusel, M., Ramsauer, H., Unterthiner, T., Nessler, B., and Hochreiter, S.
\newblock {GAN}s trained by a two time-scale update rule converge to a local
  nash equilibrium.
\newblock In \emph{Advances in Neural Information Processing Systems}, pp.\
  6626--6637, 2017.

\bibitem[Holighaus et~al.(2013)Holighaus, D{\"o}rfler, Velasco, and
  Grill]{holighaus2013framework}
Holighaus, N., D{\"o}rfler, M., Velasco, G.~A., and Grill, T.
\newblock A framework for invertible, real-time constant-{Q} transforms.
\newblock \emph{IEEE Transactions on Audio, Speech, and Language Processing},
  21\penalty0 (4):\penalty0 775--785, 2013.

\bibitem[Holighaus et~al.(2019)Holighaus, Koliander, Pr{\r{u}}{\v{s}}a, and
  Abreu]{holighaus2019char}
Holighaus, N., Koliander, G., Pr{\r{u}}{\v{s}}a, Z., and Abreu, L.~D.
\newblock Characterization of analytic wavelet transforms and a new phaseless
  reconstruction algorithm.
\newblock \emph{Preprint, submitted to IEEE Trans. Sig. Proc.}, 2019.
\newblock URL \url{http://ltfat.github.io/notes/ltfatnote053.pdf}.

\bibitem[Huang et~al.(2019)Huang, Li, Anil, Bao, Oore, and
  Grosse]{huang2018timbretron}
Huang, S., Li, Q., Anil, C., Bao, X., Oore, S., and Grosse, R.~B.
\newblock {TimbreTron: A WaveNet (CycleGAN (CQT (Audio)))} pipeline for musical
  timbre transfer.
\newblock In \emph{Proc. of ICLR}, 2019.

\bibitem[Janssen(1997)]{janssen1997continuous}
Janssen, A.
\newblock From continuous to discrete {W}eyl-{H}eisenberg frames through
  sampling.
\newblock \emph{Journal of Fourier Analysis and Applications}, 3\penalty0
  (5):\penalty0 583--596, 1997.

\bibitem[Karras et~al.(2018)Karras, Aila, Laine, and
  Lehtinen]{karras2017progressive}
Karras, T., Aila, T., Laine, S., and Lehtinen, J.
\newblock Progressive growing of {GANs} for improved quality, stability, and
  variation.
\newblock In \emph{Proc. of ICLR}, 2018.

\bibitem[Kingma \& Ba(2015)Kingma and Ba]{Kingma2014}
Kingma, D. and Ba, J.
\newblock Adam: {A} method for stochastic optimization.
\newblock In \emph{Proc. of ICLR}, 2015.

\bibitem[Le~Roux et~al.(2010)Le~Roux, Kameoka, Ono, and Sagayama]{le2010fast}
Le~Roux, J., Kameoka, H., Ono, N., and Sagayama, S.
\newblock Fast signal reconstruction from magnitude {STFT} spectrogram based on
  spectrogram consistency.
\newblock In \emph{Proc. Int. Conf. Digital Audio Effects}, volume~10, 2010.

\bibitem[Lyons(1991)]{lyons1991practical}
Lyons, L.
\newblock \emph{A Practical Guide to Data Analysis for Physical Science
  Students}.
\newblock Cambridge University Press, 1991.

\bibitem[Mallat \& Waldspurger(2015)Mallat and Waldspurger]{mallat2015phase}
Mallat, S. and Waldspurger, I.
\newblock Phase retrieval for the cauchy wavelet transform.
\newblock \emph{Journal of Fourier Analysis and Applications}, 21\penalty0
  (6):\penalty0 1251--1309, 2015.

\bibitem[Marafioti et~al.(2018)Marafioti, Perraudin, Holighaus, and
  Majdak]{marafioti2018context}
Marafioti, A., Perraudin, N., Holighaus, N., and Majdak, P.
\newblock A context encoder for audio inpainting.
\newblock \emph{Preprint, submitted to IEEE TASLP.}, 2018.
\newblock URL \url{https://arxiv.org/pdf/1810.12138.pdf}.

\bibitem[Mehri et~al.(2017)Mehri, Kumar, Gulrajani, Kumar, Jain, Sotelo,
  Courville, and Bengio]{Mehri2016}
Mehri, S., Kumar, K., Gulrajani, I., Kumar, R., Jain, S., Sotelo, J.,
  Courville, A., and Bengio, Y.
\newblock {SampleRNN}: {A}n unconditional end-to-end neural audio generation
  model.
\newblock In \emph{Proc. of ICLR}, 2017.

\bibitem[Muth et~al.(2018)Muth, Uhlich, Perraudin, Kemp, Cardinaux, and
  Mitsufuji]{muth2018improving}
Muth, J., Uhlich, S., Perraudin, N., Kemp, T., Cardinaux, F., and Mitsufuji, Y.
\newblock Improving {DNN}-based music source separation using phase features.
\newblock In \emph{Joint Workshop on Machine Learning for Music at ICML,
  IJCAI/ECAI, and AAMAS}, 2018.

\bibitem[Necciari et~al.(2018)Necciari, Holighaus, Balazs, Průša, Majdak, and
  Derrien]{necciari18}
Necciari, T., Holighaus, N., Balazs, P., Průša, Z., Majdak, P., and Derrien,
  O.
\newblock Audlet filter banks: A versatile analysis/synthesis framework using
  auditory frequency scales.
\newblock \emph{Applied Sciences}, 8\penalty0 (1:96), 2018.

\bibitem[Pascual et~al.(2017)Pascual, Bonafonte, and Serrà]{pascual2017segan}
Pascual, S., Bonafonte, A., and Serrà, J.
\newblock {SEGAN}: Speech enhancement generative adversarial network.
\newblock In \emph{Proc. of Interspeech}, pp.\  3642--3646, 2017.

\bibitem[Perraudin et~al.(2013)Perraudin, Balazs, and
  S{\o}ndergaard]{Perraudin2013}
Perraudin, N., Balazs, P., and S{\o}ndergaard, P.~L.
\newblock A fast {G}riffin-{L}im algorithm.
\newblock In \emph{Applications of Signal Processing to Audio and Acoustics
  (WASPAA), 2013 IEEE Workshop on}, pp.\  1--4. IEEE, 2013.

\bibitem[Pons et~al.(2017)Pons, Nieto, Prockup, Schmidt, Ehmann, and
  Serra]{Pons2017}
Pons, J., Nieto, O., Prockup, M., Schmidt, E.~M., Ehmann, A.~F., and Serra, X.
\newblock End-to-end learning for music audio tagging at scale.
\newblock \emph{CoRR}, abs/1711.02520, 2017.
\newblock URL \url{http://arxiv.org/abs/1711.02520}.

\bibitem[{P}ortnoff(1976)]{po76}
{P}ortnoff, M.
\newblock {I}mplementation of the digital phase vocoder using the fast fourier
  transform.
\newblock \emph{{I}{E}{E}{E} {T}rans. {A}coust. {S}peech {S}ignal {P}rocess.},
  24\penalty0 (3):\penalty0 243--248, 1976.

\bibitem[Pr\r{u}\v{s}a et~al.(2014)Pr\r{u}\v{s}a, S{\o}ndergaard, Holighaus,
  Wiesmeyr, and Balazs]{ltfatnote030}
Pr\r{u}\v{s}a, Z., S{\o}ndergaard, P.~L., Holighaus, N., Wiesmeyr, C., and
  Balazs, P.
\newblock {The Large Time-Frequency Analysis Toolbox 2.0}.
\newblock In \emph{Sound, Music, and Motion}, LNCS, pp.\  419--442. Springer
  International Publishing, 2014.
\newblock ISBN 978-3-319-12975-4.
\newblock \doi{10.1007/978-3-319-12976-1\_25}.

\bibitem[Pru{\v{s}}a(2015)]{ltfatnote042}
Pru{\v{s}}a, Z.
\newblock {STFT} and {DGT} phase conventions and phase derivatives
  interpretation.
\newblock Technical report, Acoustics Research Institute, Austrian Academy of
  Sciences, 2015.

\bibitem[Pr{\r{u}}{\v{s}}a et~al.(2017)Pr{\r{u}}{\v{s}}a, Balazs, and
  S{\o}ndergaard]{pruuvsa2017noniterative}
Pr{\r{u}}{\v{s}}a, Z., Balazs, P., and S{\o}ndergaard, P.
\newblock A noniterative method for reconstruction of phase from {STFT}
  magnitude.
\newblock \emph{IEEE/ACM Transactions on Audio, Speech and Language
  Processing}, 25\penalty0 (5):\penalty0 1154--1164, 2017.

\bibitem[Radford et~al.(2016)Radford, Metz, and
  Chintala]{radford2015unsupervised}
Radford, A., Metz, L., and Chintala, S.
\newblock Unsupervised representation learning with deep convolutional
  generative adversarial networks.
\newblock In \emph{Proc. of ICLR}, 2016.

\bibitem[Salimans et~al.(2016)Salimans, Goodfellow, Zaremba, Cheung, Radford,
  and Chen]{salimans2016improved}
Salimans, T., Goodfellow, I., Zaremba, W., Cheung, V., Radford, A., and Chen,
  X.
\newblock Improved techniques for training {GAN}s.
\newblock In \emph{Advances in Neural Information Processing Systems}, pp.\
  2234--2242, 2016.

\bibitem[Shen et~al.(2018)Shen, Pang, Weiss, Schuster, Jaitly, Yang, Chen,
  Zhang, Wang, Skerry{-}Ryan, Saurous, Agiomyrgiannakis, and
  Wu]{Tacotron2_2017}
Shen, J., Pang, R., Weiss, R., Schuster, M., Jaitly, N., Yang, Z., Chen, Z.,
  Zhang, Y., Wang, Y., Skerry{-}Ryan, R., Saurous, R., Agiomyrgiannakis, Y.,
  and Wu, Y.
\newblock Natural {TTS} synthesis by conditioning {WaveNet} on mel spectrogram
  predictions.
\newblock In \emph{Proc. of ICASSP}, 2018.

\bibitem[Sotelo et~al.(2017)Sotelo, Mehri, Kumar, Santos, Kastner, Courville,
  and Bengio]{sotelo2017char2wav}
Sotelo, J., Mehri, S., Kumar, K., Santos, J.~F., Kastner, K., Courville, A.,
  and Bengio, Y.
\newblock Char2wav: {E}nd-to-end speech synthesis.
\newblock In \emph{Workshop on ICLR}, 2017.

\bibitem[{S}trohmer(1998)]{st98-1}
{S}trohmer, T.
\newblock {N}umerical algorithms for discrete {G}abor expansions.
\newblock In {F}eichtinger, H.~G. and {S}trohmer, T. (eds.), \emph{{G}abor
  {A}nalysis and {A}lgorithms: {T}heory and {A}pplications}, {A}ppl. {N}umer.
  {H}armon. {A}nal., pp.\  267--294. {B}irkh{\"a}user {B}oston, 1998.

\bibitem[{S}trohmer \& {B}eaver(2003){S}trohmer and {B}eaver]{best03}
{S}trohmer, T. and {B}eaver, S.
\newblock {O}ptimal {O}{F}{D}{M} system design for time-frequency dispersive
  channels.
\newblock \emph{{I}{E}{E}{E} {T}rans. {C}omm.}, 51\penalty0 (7):\penalty0
  1111--1122, {J}uly 2003.

\bibitem[Van Den~Oord et~al.(2016)Van Den~Oord, Dieleman, Zen, Simonyan,
  Vinyals, Graves, Kalchbrenner, Senior, and Kavukcuoglu]{Wavenet2016}
Van Den~Oord, A., Dieleman, S., Zen, H., Simonyan, K., Vinyals, O., Graves, A.,
  Kalchbrenner, N., Senior, A., and Kavukcuoglu, K.
\newblock Wavenet: A generative model for raw audio.
\newblock \emph{CoRR abs/1609.03499}, 2016.

\bibitem[Van Den~Oord et~al.(2018)Van Den~Oord, Li, Babuschkin, Simonyan,
  Vinyals, Kavukcuoglu, van~den Driessche, Lockhart, Cobo, Stimberg,
  Casagrande, Grewe, Noury, Dieleman, Elsen, Kalchbrenner, Zen, Graves, King,
  Walters, Belov, and Hassabis]{parallel_wavenet2017}
Van Den~Oord, A., Li, Y., Babuschkin, I., Simonyan, K., Vinyals, O.,
  Kavukcuoglu, K., van~den Driessche, G., Lockhart, E., Cobo, L., Stimberg, F.,
  Casagrande, N., Grewe, D., Noury, S., Dieleman, S., Elsen, E., Kalchbrenner,
  N., Zen, H., Graves, A., King, H., Walters, T., Belov, D., and Hassabis, D.
\newblock Parallel {W}ave{N}et: Fast high-fidelity speech synthesis.
\newblock In \emph{Proc. of ICML}, pp.\  3918--3926, 2018.

\bibitem[Warden(2018)]{warden2018speech}
Warden, P.
\newblock Speech commands: A dataset for limited-vocabulary speech recognition.
\newblock \emph{arXiv preprint arXiv:1804.03209}, 2018.

\bibitem[Wexler \& Raz(1990)Wexler and Raz]{rawe90}
Wexler, J. and Raz, S.
\newblock Discrete gabor expansions.
\newblock \emph{Signal Processing}, 21\penalty0 (3):\penalty0 207 -- 220, 1990.
\newblock \doi{https://doi.org/10.1016/0165-1684(90)90087-F}.

\end{thebibliography}
\bibliographystyle{icml2019}

\appendix
\newpage
\twocolumn[\huge{\centering \textbf{Supplementary Material}\\ \vspace{1cm} }]

\section{Detail of the the GAN architecture}\label{sec:supp-gandetails}

Table \ref{tab:gan_architecture} presents the details of the convolutional architecture used by TFGAN-M and TFGAN-MTF. Here  $B = 64$ is the batch size. 

\begin{table}[ht]
\centering
\begin{small}
\begin{tabular}{|l|l|l|}
 \hline
  Operation & Kernel Size& Output Shape  \\
 \hline
\multicolumn{3}{c}{Generator} \\
\hline
Input $z ~ \mathcal{N}(0,1)$ & & $(B,100)$\\
Dense & $(100, 256s)$ & $(B, 256s)$\\
Reshape & & $(B,8,4,8s)$ \\ 
DeConv 2D (Stride 2) & $(12,3,8s,8s)$ & $(B, 16, 8, 8s)$\\
LReLu $(\alpha=0.2)$  &  & $(B, 16, 8, 8s)$\\
DeConv 2D (Stride 2) & $(12,3,8s,4s)$ & $(B, 32s, 16, 4s)$\\
LReLu $(\alpha=0.2)$  &  & $(B, 32s, 16, 4s)$\\
DeConv 2D (Stride 2) & $(12,3,4s,2s)$ & $(B, 64, 32, 2s)$\\
LReLu $(\alpha=0.2)$  &  & $(B, 64, 32, 2s)$\\
DeConv 2D (Stride 2) & $(12,3,2s,s)$ & $(B, 128, 64, s)$\\
LReLu $(\alpha=0.2)$  &  & $(B, 32, 32, 2d)$\\
DeConv 2D (Stride 2) & $(12,3,s,c)$ & $(B, 256, 128,  c)$\\
\hline
\multicolumn{3}{c}{Discriminator} \\
\hline
Input  & & $(B, 256, 128,  c)$\\
Conv 2D (Stride 2) & $(12,3,c,s) $& $(B, 128, 64,  s)$\\
LReLu $(\alpha=0.2)$  &  & $(B, 128, 64,  s)$\\
Conv 2D (Stride 2) & $(12,3,s,2s)$ & $(B, 64, 32,  2s)$\\
LReLu $(\alpha=0.2)$  &  & $(B, 64, 32,  2s)$\\
Conv 2D (Stride 2) & $(12,3,2s,4s)$ & $(B, 32, 16,  4s)$\\
LReLu $(\alpha=0.2)$  &  & $(B, 32, 16,  4s)$\\
Conv 2D (Stride 2) & $(12,3,4s,8s)$ & $(B, 16, 8,  8s)$\\
LReLu $(\alpha=0.2)$  &  & $(B, 16, 8,  8s)$\\
Conv 2D (Stride 2) & $(12,3,8s,16s)$ & $(B, 8, 4,  16s)$\\
LReLu $(\alpha=0.2)$  &  & $(B, 8, 4,  16s)$\\
Reshape &  & $(B, 512s)$\\
Dense & $(512s, 1)$ & $(B,1)$ \\
\hline
\end{tabular}
\caption{Detailed architecture of the Generative adversarial network. Scale $s=64$. Channels $c=1$ for the magnitude network and $c=3$ for the network that also outputs the derivatives.
\label{tab:gan_architecture}
}
\end{small}
\end{table}


\section{Listening test location}\label{sec:supp-booths}
\begin{figure}[h]
       \centering
       \includegraphics[width=.49\textwidth]{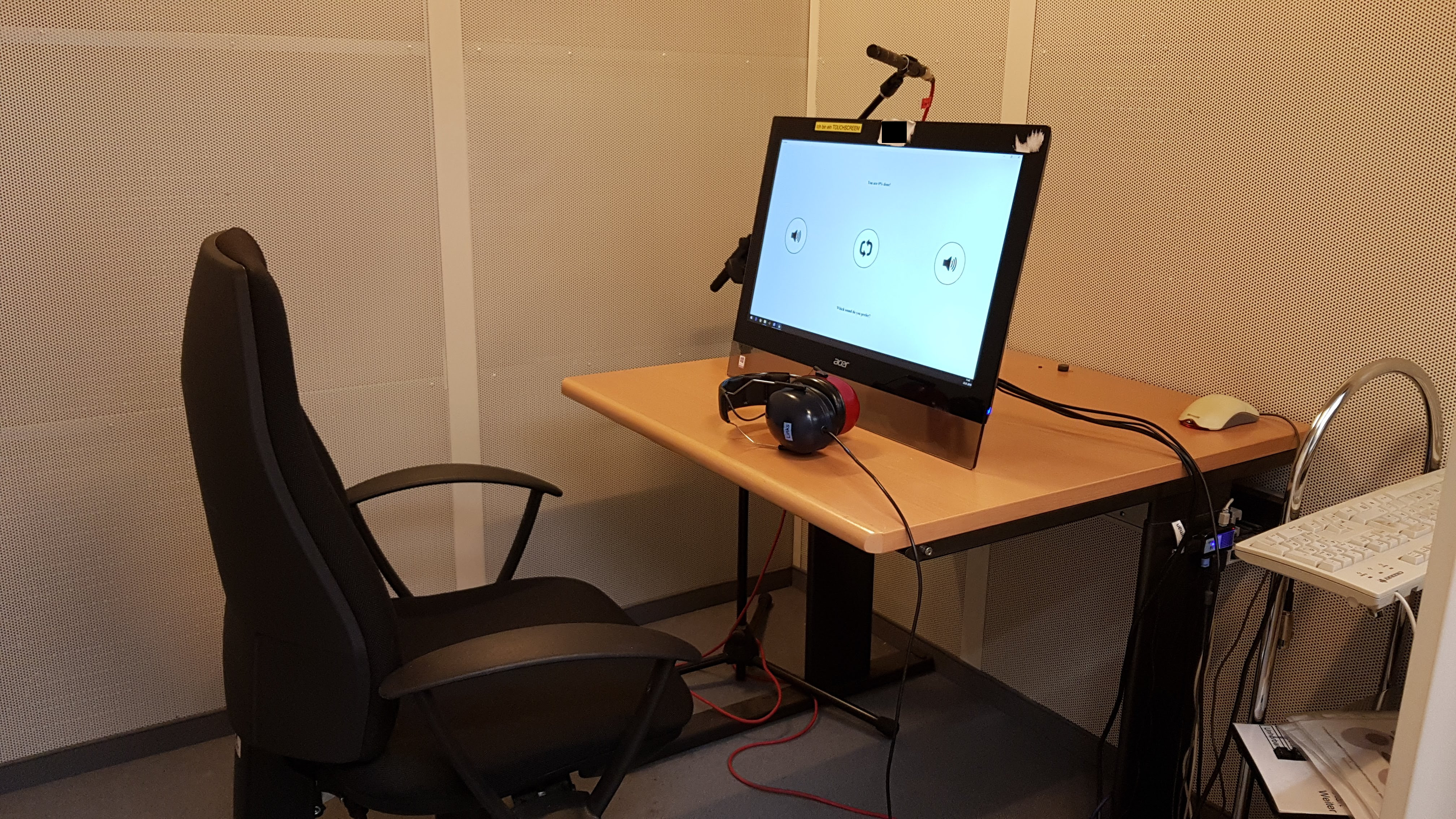}
       \caption{Inside of the sound booth used to perform the listening test.}
      \label{fig:booth}
   \end{figure}

\begin{figure}[h]
       \centering
       \includegraphics[width=.49\textwidth]{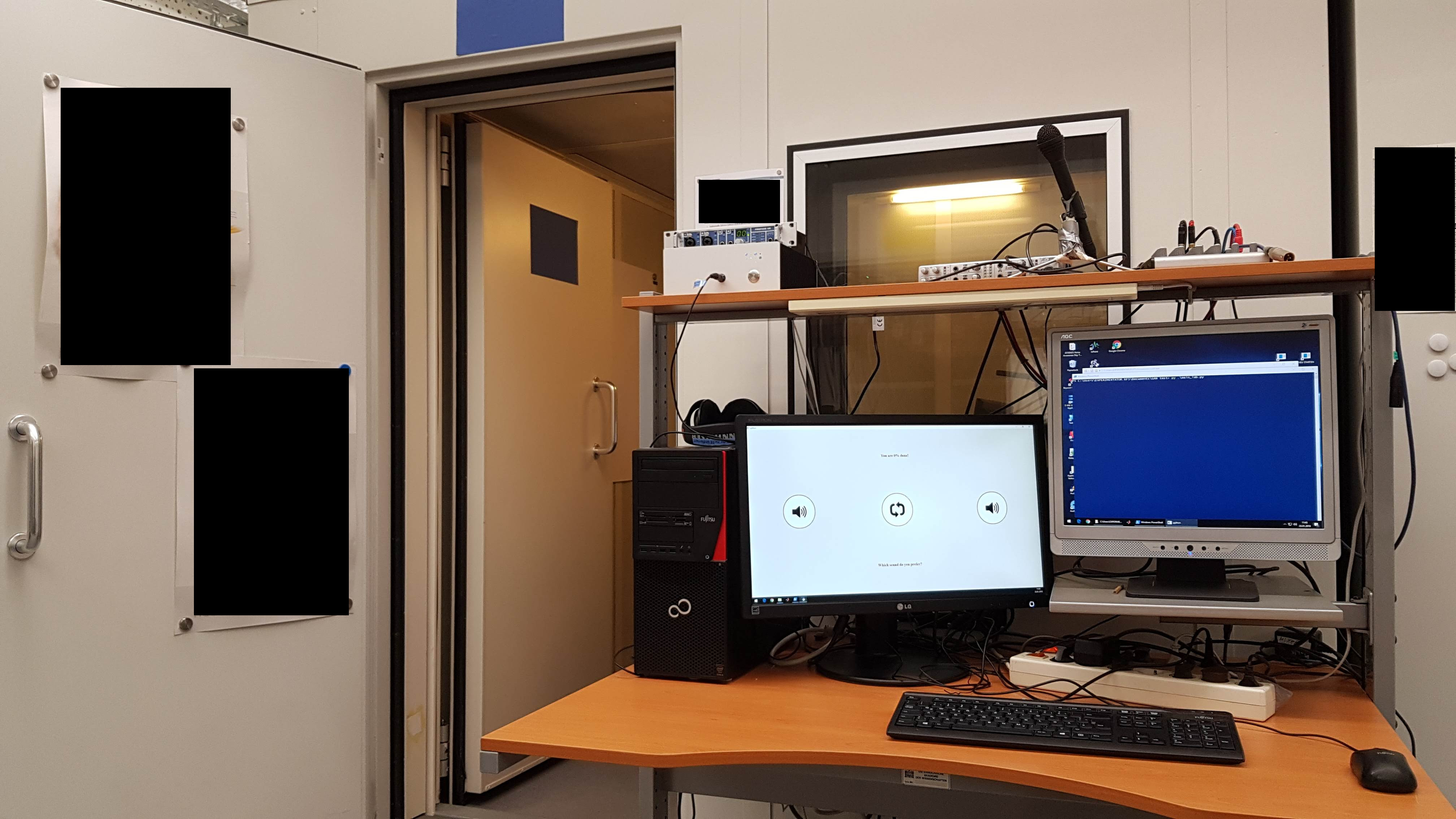}
       \caption{Sound booth from the outside with equipment for external monitoring of ongoing tests.}
      \label{fig:control}
   \end{figure}

 Figures \ref{fig:booth} and \ref{fig:control} show the physical setup of the listening test, including the sound booth and additional equipment.

\section{Comparison to GANSynth:} A direct comparison between the results of the very recent GANSynth architecture~ \cite{engel2019gansynth}, which obtained unprecedented audio quality for adversarial audio synthesis, and TFGAN 
is not straightforward, since GANSynth considers semi-supervised generation conditioned on pitch, while we considered unsupervised generation to facilitate the comparison with WaveGAN. Further, the network architecture \cite{karras2017progressive} on which GANSynth is built is significantly larger and more sophisticated than our DCGAN-derived architecture. The adaptation of TFGAN to a comparable architecture and its application to semi-supervised generation generation is planned for future work. For now, we can only observe that a key ingredient of GANSynth is the usage of the time-direction phase derivative, which in fact
corroborates our claim that careful modeling of the structure of the STFT is crucial for neural generation of time-frequency features.
As discussed in Section \ref{ssec:phaserec}, the PLR method employed in GANSynth can be unreliable and synthesis quality can likely be further improved if a more robust method for PLR is considered. Audio examples for different PLR methods are provided in the supplementary material.
   
\section{Short-time Fourier phase conventions}\label{sec:supp-conventions}

  In Section \ref{sec:discSTFT}, we introduced the STFT in the so-called \emph{frequency-invariant} convention. This is the convention preferred in the mathematical community. It arises from the formulation of the discrete STFT as a sliding window DFT. There are various other conventions, depending on how the STFT is derived and implemented. Usually, the chosen convention does not affect the magnitude, but only the phase of the STFT. When phase information is processed, it is crucial to be aware of the convention in which the STFT is computed, and adapt the processing scheme accordingly.
  Usually, the conversion between conventions amounts to the point-wise multiplication of the STFT with a predetermined matrix of phase factors. Common \emph{phase conventions} and the conversion between them are discussed in \cite{dolson1986phase,arfib2011time}. The $3$ most wide-spread conventions, the last of which is rarely described, but frequently implemented in software frameworks, are presented here:
  
  \paragraph{Frequency-invariant STFT:} The $m$-th channel of the frequency-invariant STFT can be interpreted as demodulating the signal $s$ with the pure frequency $e^{-2\pi i ml/M}$, before applying a low-pass filter with impulse response $g[-\cdot]$. Therefore, the phase is expected to change slowly in the time-direction, it is already demodulated. The time-direction phase derivative indicates the distance of the current position to the local instantaneous frequency. On the other hand, the evolution of the phase in frequency-direction depends on the time position. Hence, the frequency-direction derivative indicates the (absolute) local group delay, sometimes called the instantaneous time.  
  
  \paragraph{Time-invariant STFT:} Given by
       \begin{equation}\label{eq:STFT_tinv}
            \begin{split}
            \operatorname{STFT}^{\textrm{ti}}_{g}(s)[m,n] \\
             & = \sum_{l=-\lfloor L_g/2 \rfloor}^{\lceil L_g/2 \rceil-1} s[l+na]g[l]e^{-2\pi i ml/M},
         \end{split}
         \end{equation}
     the time-invariant STFT can be interpreted as filtering the signal $s$ with the band-pass filters $g[-\cdot]e^{2\pi i m(\cdot)/M}$. Hence, the phase is expected to revolve at roughly the channel center frequency in the time-direction and the time-direction phase derivative points to the (absolute) local instantaneous frequency. In the frequency direction, however, the phase in the frequency-direction changes slowly, i.e. it is demodulated in the frequency-direction. The frequency-direction phase derivative indicates the distance to the local instantaneous time. 
In each, the frequency- and time-invariant STFT, the phase is demodulated in one direction, but moves quickly in the other. In Section \ref{ssec:phaserec}, we propose to use the derivative of the demodulated phase in both directions, such that we must convert between the two conventions. This conversion is achieved simply by pointwise multiplication of the STFT matrix with a matrix of phase factors:
\begin{equation}\label{eq:phaseconversionFITI}
  \begin{split}
  \operatorname{STFT}_{g}(s)[m,n] & = e^{-2\pi i mna/M}\operatorname{STFT}^{\textrm{ti}}_{g}(s)[m,n] \\
  & = W[m,n]\operatorname{STFT}^{\textrm{ti}}_{g}(s)[m,n].  
  \end{split}
\end{equation}
  Equivalently, if $\phi_g = \arg(\operatorname{STFT}_{g}(s))$ is the phase of the frequency-invariant STFT and $\phi_g^{\textrm{ti}}=\arg(\operatorname{STFT}^{\textrm{ti}}_{g}(s))$, then $\phi_g[m,n] = \phi_g^{\textrm{ti}}[m,n] -2\pi i mna/M$.
       
  \paragraph{Simplified time-invariant STFT:} In many common frameworks, including SciPy and Tensorflow, the STFT computation follows neither the frequency- nor time-invariant conventions. Instead, the window $g$ is stored as a 
  vector of length $L_g$ with the peak not at $g[0]$, but at $g[\lfloor L_g/2 \rfloor]$. The STFT is then computed as 
  \begin{equation}\label{eq:STFT_stinv}
            \begin{split}
            \operatorname{STFT}^{\textrm{sti}}_{g}(s)[m,n]
             & = \sum_{l=0}^{L_g-1} s[l+na]g[l]e^{-2\pi i ml/M}.
         \end{split}
   \end{equation}
   The above equation is slightly easier to implement, compared to the frequency- or time-invariant STFT, if $M\geq L_g$, since in that case, $g\in\RR^{L_g}$ can simply be zero-extended to length $M$, after which the following holds: $\operatorname{STFT}^{\textrm{sti}}_{g}(s)[\cdot,n] = \textrm{DFT}_M(s^{(n)})[m]$, with $s^{(n)} = [s[na]g[0],\ldots,s[na+M-1]g[M-1]]^T \in \RR^M$. Comparing \eqref{eq:STFT_tinv} with \eqref{eq:STFT_stinv}, we can see that the latter introduces a delay and a phase skew dependent on the (stored) window length $L_g$. In general, we obtain the equality 
   \begin{equation}
     \begin{split}
       & \operatorname{STFT}^{\textrm{sti}}_{g}(s)[m,n] \\
       & = e^{-2\pi i m\lfloor L_g/2 \rfloor/M}\operatorname{STFT}^{\textrm{ti}}_{g}(s[\cdot+\lfloor L_g/2 \rfloor])[m,n].
     \end{split}
   \end{equation}
   If the hop size $a$ is a divisor of $\lfloor L_g/2 \rfloor$, then we can convert into a time-invariant STFT: 
   \begin{equation}\label{eq:phaseconversionSTITI}
     \begin{split}
        & \operatorname{STFT}^{\textrm{ti}}_{g}(s)[m,n+\lfloor L_g/2 \rfloor/a] \\
        & = e^{2\pi i m\lfloor L_g/2 \rfloor/M}\operatorname{STFT}^{\textrm{sti}}_{g}(s)[m,n],  
   \end{split}
   \end{equation}
   or equivalently $\phi_g[m,n+\lfloor L_g/2 \rfloor/a] = \phi_g^{\textrm{ti}}[m,n+\lfloor L_g/2 \rfloor/a] -2\pi i m(na+\lfloor L_g/2 \rfloor)/M = \phi_g^{\textrm{sti}}[m,n] -2\pi i mna/M$. Note that, additionally, SciPy and Tensorflow do not consider $s$ circularly, but compute only those time frames, for which the window does not overlap the signal borders, i.e., $n\in[0,\ldots,\lfloor(L-L_g)/a\rfloor]$. If an STFT according to the convention \eqref{eq:STFT_stinv}, with $N$ time frames and aligned with the time-invariant STFT is desired, the signal $s$ can be extended to length $L+L_g$ by adding $\lfloor L_g/2 \rfloor$ zeros before $s[0]$ and $\lceil L_g/2 \rceil$ zeros after $s[L-1]$. 
%
%

\end{document}